\documentclass[aps,reprint,superscriptaddress,floatfix,nofootinbib]{revtex4-2}
\usepackage{amsmath,amssymb,graphicx,booktabs}
\usepackage{subcaption}
\DeclareCaptionJustification{flush}{\flushing}
\usepackage{xcolor}
\definecolor{linkblue}{RGB}{28,84,148}

\usepackage[colorlinks=true,linkcolor=linkblue,citecolor=linkblue,urlcolor=linkblue]{hyperref}
\usepackage[capitalise,nameinlink]{cleveref}
\crefname{figure}{Fig.}{Figs.}
\Crefname{figure}{Figure}{Figures}
\crefname{equation}{Eq.}{Eqs.}
\Crefname{equation}{Equation}{Equations}
\crefname{section}{Sec.}{Secs.}
\crefname{table}{Table}{Tables}
\hypersetup{pdftitle={Reaching high fusion gain with grams of spin-polarized fuel},pdfauthor={J. F. Parisi}}
\begin{document}
\title{Reaching high fusion gain with grams of spin-polarized fuel}
\author{J. F. Parisi}
\email{jason@marathonfusion.com}
\affiliation{Marathon Fusion, 150 Mississippi Street, San Francisco, CA 94107, USA}
\begin{abstract}
Building on recent ignition-access work of Delgado-Aparicio, Ono, and Menard, we show that even a single-use, gram-scale quantity of spin-polarized fuel (SPF) is useful for increasing fusion power and gain in magnetic confinement fusion machines such as tokamaks and stellarators. While fueling a fusion power plant continuously with SPF requires $\sim$kilograms per day, far beyond present capabilities, a single-use short pulse of SPF allows a fusion plasma to cross into a high-gain regime, and stay there, even after switching back to regular unpolarized fuel. With continuous polarized fueling, polarization also makes high-gain plasmas easier to control: a resonant wave that depolarizes the fuel lowers the reactivity quickly, on a much faster timescale than transport. Grams of polarized fuel can therefore improve plasma performance and open a path to scaling SPF sources and usage, from small first experiments to continuous fueling.
\end{abstract}
\maketitle

\section{Introduction}

A net electricity-producing fusion power plant must typically produce far more power than it spends on heating its plasma~\cite{lawson,wurzel2022}. For many confinement schemes such as tokamaks~\cite{shimada2007,creely2020,hillesheim2026overview,meyer2024,willis2026,manta2024,brown2026,siccinio2020,zhuang2019,kim2015,tobita2019} and stellarators~\cite{lion2025,hegna2025,swanson2026,lazerson2026,prost2024,helias,sagara2014}, this requires a high-gain plasma, heated mainly by alpha particles from deuterium-tritium (D-T) fusion, which can be hard to access, hard to control, and slow to exit. High-gain access is uncertain because the required auxiliary heating depends on confinement not yet tested in tokamak or stellarator burning plasmas~\cite{ipb98,verdoolaege2021} and contingencies such as a larger device and spare heating capacity are expensive~\cite{wade2021}. Control is difficult because an ignited plasma --- which sustains itself without auxiliary heating --- no longer has heating as a significant control knob, and the remaining actuators (fueling and impurities) work only through the density and temperature~\cite{schuster2003,boyer2015,humphreys2015}. A controlled reduction of the fusion power (after a fault or in a thermal runaway, an increase in temperature that feeds itself through the alpha heating~\cite{schuster2003,boyer2015}) therefore takes particle and energy confinement timescales~\cite{strait,devries2018}, and the faster alternative, massive impurity injection~\cite{fpss}, ends the discharge in a mitigated disruption, with loads on the first wall that a power plant may not tolerate~\cite{siccinio2020}. 

Polarizing the nuclear spins of the fuel increases the D-T fusion cross section by up to 50\%~\cite{kulsrud,cowley1986,hupin2019}, and can help with all three of these challenges. In particular, we will argue that even gram-scale quantities of spin-polarized fuel (SPF) are useful, by lowering the heating needed for high-gain access and with it the science risk and the installed heating capacity. With continuous polarized fueling, an intentionally launched depolarizing wave could lower the cross section rapidly, and is a possible fast means of control and exit that can also stop thermal runaways.

SPF for nuclear fusion has been studied since the 1980s~\cite{kulsrud,more1983,coppi1983,micklich1984,kvc,cowley1986,paetz2010}. Recent analytic and numerical work has calculated the spin-polarized cross section from first principles~\cite{hupin2019}, and studied how the polarization survives inertial-fusion implosions~\cite{hu2023} and plasma waves~\cite{cook}. Notably, the first experimental observation that nuclear polarization survives in a laser-driven plasma was recently reported~\cite{zheng2026}. Others are developing ways to produce SPF at scale~\cite{ciullo2016,spiliotis2021,kannis2021,ciullo2022,kannis2026} and are designing experiments to measure its lifetime in the DIII-D tokamak~\cite{baylor,garcia2023,heidbrink2024,garcia2025}. In addition to the well-known benefit of SPF increasing fusion reactivity, recent analytic and numerical work has shown how SPF benefits a fusion power plant in tritium burn efficiency~\cite{parisi2024}, neutronics~\cite{bae2025}, net electric power~\cite{parisi2025power}, D-$^3$He fusion~\cite{parisi2025dhe3} and alpha channeling~\cite{parisi2026channel}.

\begin{figure}[b]
\centering
\includegraphics[width=\linewidth]{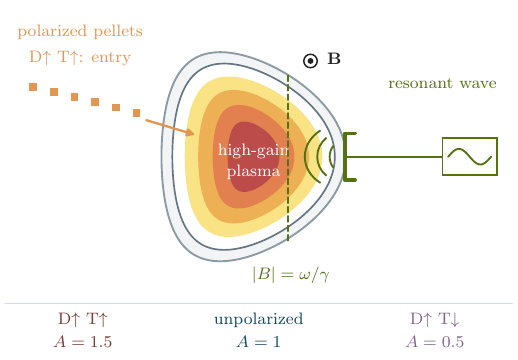}
\caption{High-gain access, egress, and control with spin-polarized fuel. Polarized fuel, injected here as pellets, increases the fusion reactivity on the approach to high gain. A wave of angular frequency $\omega$, resonant where field strength $|B|=\omega/\gamma$ for nuclei of gyromagnetic ratio $\gamma$, depolarizes.}
\label{fig:schematic}
\end{figure}

\begin{figure*}[t]
\centering
\begin{subfigure}[t]{0.495\textwidth}
\includegraphics[width=\linewidth]{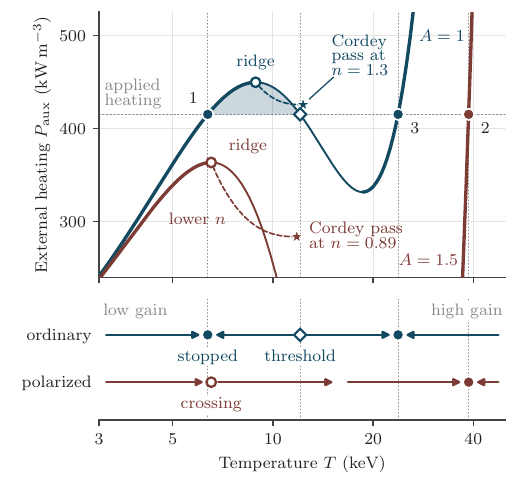}
\caption{}\label{fig:states-a}
\end{subfigure}\hfill
\begin{subfigure}[t]{0.495\textwidth}
\includegraphics[width=\linewidth]{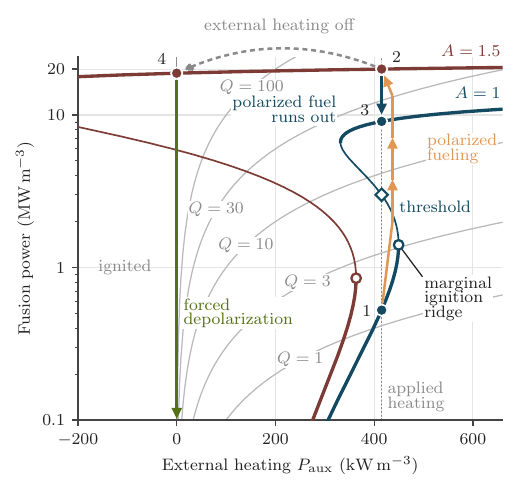}
\caption{}\label{fig:states-b}
\end{subfigure}
\caption{Steady states for $A=1$ and 1.5 in the radiating plasma at fuel density $n=1.55\times10^{20}$ m$^{-3}$. Here, only polarized fuel reaches high gain. Thin lines are thermally unstable states, where required heating decreases as temperature increases. (\subref{fig:states-a}) External heating needed for a steady state, $P_{\rm aux}$ versus temperature (top) and direction of temperature change with applied heating (bottom). The plasma cools where curve is above the applied heating (shaded) and heats when below: ordinary fuel stops at state 1 and polarized fuel heats to state 2. Open circles: marginal ignition ridge at this density. Dashed lines follow the ridge as density decreases, down to its lowest point, the Cordey pass (stars, labeled with fuel density in $10^{20}$ m$^{-3}$), which a plasma heated at low density and fueled afterwards must cross. (\subref{fig:states-b}) Fusion power of the same states versus heating. Diamond: unstable state that ordinary fuel must pass to reach high gain. Sequence: crossing into high gain with polarized fuel (1 to 2), polarized fuel running out (2 to 3), external heating off (2 to 4) and forced depolarization.}
\label{fig:states}
\end{figure*}

Lawson was the first to specify the conditions required for ignition (density $n$, temperature $T$, and energy confinement time $\tau_\mathrm{E}$)~\cite{lawson}, but said little of the pathway to it, such as the external heating required. Polarized fuel has long been known to ease ignition access, in tokamaks~\cite{micklich1984,mitarai} and in inertial fusion~\cite{more1983,temporal2012}. Mitarai, Hasuyama, and Wakuta found that complete polarization reduces the least heating needed to reach ignition by 17 to 24\% in two D-T tokamak designs, and proposed using polarization only transiently, with polarized pellets during startup, to reduce the auxiliary heating~\cite{mitarai}. In this work we calculate how much polarized fuel this needs and show that the plasma stays at high gain on ordinary fuel after the polarized fuel has been replaced.

Delgado-Aparicio, Ono, and Menard recently described ignition access through the auxiliary heating $P_{\rm aux}(n,T)$ that keeps a plasma of fuel density $n$ at a constant temperature $T$~\cite{lfdav}. The plasma is ignited where $P_{\rm aux}\leq0$. The maxima of $P_{\rm aux}$ over temperature at each density form a marginal ignition ridge, whose lowest point, the Cordey pass~\cite{mitarai}, is the least heating with which a plasma heated first and fueled afterwards can reach high gain. Beyond the ridge the alpha heating plus the applied heating exceeds the losses, and the plasma heats itself onto a hot branch with high gain, where the remaining external heating is only a few percent of the fusion power [\cref{fig:states}(a)]. With less heating than the ridge needs, the plasma stays on the cool side at low gain.

\begin{figure*}[t]
\centering
\begin{subfigure}[t]{0.325\textwidth}
\includegraphics[width=\linewidth]{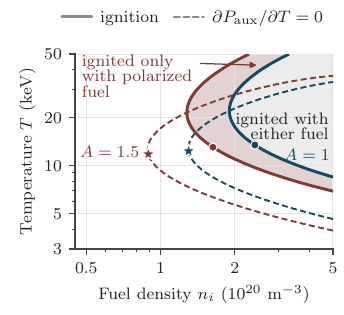}
\caption{Ignition boundaries}\label{fig:access-a}
\end{subfigure}\hfill
\begin{subfigure}[t]{0.325\textwidth}
\includegraphics[width=\linewidth]{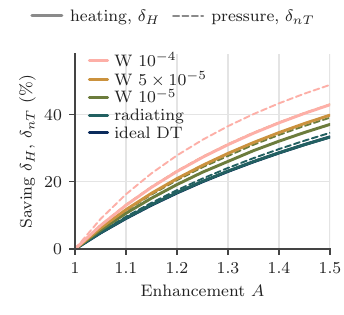}
\caption{Cordey pass and entry point}\label{fig:access-b}
\end{subfigure}\hfill
\begin{subfigure}[t]{0.325\textwidth}
\includegraphics[width=\linewidth]{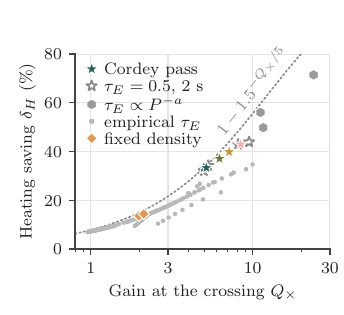}
\caption{Heating saving and crossing gain}\label{fig:access-c}
\end{subfigure}
\caption{Ignition access at $\tau_E=1$ s, computed with power balance in Ref.~\cite{lfdav}. (\subref{fig:access-a}) Ignition boundaries (solid) and $\partial P_{\rm aux}/\partial T=0$ (dashed, the marginal ignition ridge below its leftmost point) for $A=1$ and 1.5 in the radiating plasma. The plasma is ignited to the right of each solid line: with either fuel in the grey region, and only with polarized fuel in colored band. Stars: Cordey passes. Circles: entry points. (\subref{fig:access-b}) Heating saving $\delta_H$ at the Cordey pass (solid) and pressure saving $\delta_{nT}$ at the entry point (dashed) for the cases of \cref{tab:cases}. (\subref{fig:access-c}) Heating saving $\delta_H$ at $A=1.5$ versus gain at the crossing $Q_\times$.}
\label{fig:access}
\end{figure*}

Delgado-Aparicio, Ono, and Menard also showed that polarized fuel reopens the ignited region that is narrowed by radiation, roughly halving the pressure at the Cordey pass with 100 ppm of tungsten~\cite{lfdav}. We build directly on their work: with the same power balance and cases, we reproduce this result and add a rule for the external heating saved, time-dependent simulations of the approach to high gain, and an exit from ignition (\cref{fig:schematic,fig:states}). The heating needed to cross into high gain falls with the cross-section enhancement $A$ at the rate $P_\alpha/A$, with $P_\alpha$ the alpha heating at the crossing, so the heating saving increases with the fusion gain there (\cref{fig:access}). Polarized fuel is needed only during the crossing (\cref{fig:path}): grams, whereas continuous polarized fueling needs about 6 kg per day per gigawatt of fusion power at a tritium burn efficiency of 4\%~\cite{abdou,whyte2023}. Some plasmas ignite with polarized fuel but would not with ordinary fuel, and in these a wave that depolarizes the fuel can end ignition rapidly (\cref{fig:exit}).

\section{Models}\label{sec:model}

We first describe the models used in this work. For a lot of the formalism, we adopt the treatment of Delgado-Aparicio, Ono, and Menard~\cite{lfdav}. A plasma with fuel ion density $n$, a common electron and ion temperature $T$, and thermal energy $W=3nT$ loses energy by transport at the rate $W/\tau_E$, with energy confinement time $\tau_E=1$ s, and by bremsstrahlung, line and synchrotron radiation in a 10 T field. Helium ash builds up to the density $S\rho\tau_E$, where $S$ is the volumetric fusion rate and $\rho=5$ the ratio of the helium and energy confinement times, and impurities have fixed concentrations relative to the electron density $n_e$. The auxiliary heating in steady-state power balance is
\begin{equation}
P_{\rm aux}(n,T)=\frac{W}{\tau_E}+P_{\rm rad}+P_{\rm syn}-P_\alpha,
\label{eq:paux}
\end{equation}
with $P_{\rm rad}$ the radiated power, $P_{\rm syn}$ the synchrotron power and $P_\alpha$ the alpha heating, all per unit volume. We use the five cases of Ref.~\cite{lfdav}. The first, which we call ideal, is a pure D-T plasma that loses energy only by transport, with no helium ash and no radiation. The radiating plasma adds ash, bremsstrahlung and synchrotron radiation, and three further cases add 1\% nitrogen, 0.1\% argon and 10, 50 or 100 ppm of tungsten. Polarization multiplies the fusion cross section by a factor $A$, which modifies the cross section as follows,
\begin{equation}
 \sigma_{\rm pol}(E)=A\,\sigma(E),\qquad A=1+\frac{p_\mathrm{D}p_\mathrm{T}}{2},
\label{eq:A}
\end{equation}
where $\sigma(E)$ and $\sigma_{\rm pol}(E)$ are the fusion cross sections of ordinary and polarized fuel at center-of-mass energy $E$, $p_\mathrm{D}=(N_\mathrm{D}^+-N_\mathrm{D}^-)/N_\mathrm{D}$ is the vector polarization of the deuterons, with $N_\mathrm{D}^\pm$ the densities of deuterons with spin projection $\pm1$ and $N_\mathrm{D}$ their total density, and $p_\mathrm{T}=(N_\mathrm{T}^+-N_\mathrm{T}^-)/N_\mathrm{T}$ is that of the tritons, with spin projections $\pm1/2$~\cite{kulsrud,hupin2019}. $A$ is energy-independent so at a given temperature polarization multiplies the reactivity $\langle\sigma v\rangle$ by $A$. Any further fusion power increase comes from the higher temperature in a polarized plasma (which our model includes). Vector-aligned fuel ($p_\mathrm{D}=p_\mathrm{T}=1$) gives $A=1.5$, and fuel with opposed spins ($p_\mathrm{D}=\pm1$, $p_\mathrm{T}=\mp1$) gives $A=0.5$.

We use a time-dependent model of the radiating plasma with the same $\tau_E$ and field and a helium particle confinement time of 5 s. Our model contains the effect of alpha slowing and evolves the spin populations of the deuterons and tritons. Fuel is replaced on a time $\tau_f$ by new fuel whose polarization can differ between the isotopes, fuel with aligned spins fuses preferentially~\cite{takahashi}, and a depolarizing wave enters as a relaxation rate for the spins of one species (not both, because the nuclear precession frequencies of deuterons and tritons differ by a factor of 6.9). A fuel ion is either replaced, after the time $\tau_f$ (its particle confinement time, since the model has no recycling), or fuses, after a mean time $\tau_{\rm fus}$ of 13 to 19 s in our plasmas, so the tritium burn efficiency (TBE)~\cite{whyte2023} is $\tau_f/(\tau_f+\tau_{\rm fus})$. We use $\tau_f=2$ s, corresponding to a TBE of 10 to 13\%, and vary $\tau_f$ from 0.25 to 50 s. With ordinary (unpolarized) fuel the time-dependent model needs 12\% more heating to cross into high gain than \cref{eq:paux} -- this difference is because it also includes the thermal energy of the helium and its electrons, the heating of new fuel, and the fuel displaced by fast alphas.

\section{High-gain access}\label{sec:access}

High-gain access depends on two main requirements: the external heating needed to cross the ridge, which fixes the required heating capacity to install, and the plasma pressure where ignition begins. In this section we calculate how much polarization lowers both the required heating and pressure: we derive a simple rule for the heating and pressure and then test both in the five cases of \cref{sec:model} (\cref{fig:access}). 

We define $H_\times$ as the heating at which the plasma crosses the ridge onto the high-gain branch: at the Cordey pass, a saddle point of $P_{\rm aux}(n,T)$ where $\partial P_{\rm aux}/\partial n=\partial P_{\rm aux}/\partial T=0$, or, for a plasma heated at fixed density, the cool-side maximum, where $\partial P_{\rm aux}/\partial T=0$. When $A$ changes, the crossing point moves, but $H_\times$ is unaffected to first order because $P_{\rm aux}$ has zero derivative in each variable. Note this is still true regardless of how the saddle is oriented. With strong tungsten radiation, the Cordey pass is actually oriented the other way: the heating needed to keep the plasma hot is highest at the required density of the Cordey pass. So a plasma heated first and fueled afterwards needs at least this much heating to stay hot as its density increases~\cite{lfdav}. Thus $dH_\times/dA=\partial P_{\rm aux}/\partial A$ at fixed $n$ and $T$, and, neglecting the extra helium ash that the higher reactivity produces,
\begin{equation}
 -\frac{dH_\times}{dA}=\frac{P_{\alpha\times}}{A},
\label{eq:rule}
\end{equation}
where $P_{\alpha\times}$ is the alpha heating at the crossing. With the fusion power $P_f=5P_\alpha$ and the gain $Q_\times=P_f/H_\times$, \cref{eq:rule} becomes
\begin{equation}
\frac{d \ln H_\times}{d \ln A} = -\frac{Q_\times}{5}.
\label{eq:gainrule}
\end{equation}
We define the heating saving $\delta_H$ as the fraction of the crossing heating that is removed by polarization,
\begin{equation}
 \delta_H\equiv1-\frac{H_\times(A)}{H_\times(1)}\simeq1-A^{-Q_\times/5},
\label{eq:saving}
\end{equation}
where the third term is obtained by integrating \cref{eq:gainrule} at constant $Q_\times$, evaluated at $A=1$. We call \cref{eq:saving} the heating rule. Because $Q_\times$ decreases as $A$ increases in every case we considered here, \cref{eq:saving} is an upper bound on $\delta_H$. For fully polarized fuel, $\delta_H$ is 15\% at $Q_\times=2$, 33\% at $Q_\times=5$ and 50\% at $Q_\times=8.5$. On the ignition boundary the generalized Lawson criterion of Ref.~\cite{lfdav} can be written as
\begin{equation}
 nT\tau_E=\frac{12T^2}{(A-R)\langle\sigma v\rangle E_\alpha},\qquad R=\frac{P_{\rm rad}+P_{\rm syn}}{P_\alpha^{(1)}},
\label{eq:lawson}
\end{equation}
where $E_\alpha=3.5$ MeV is the alpha birth energy, $P_\alpha^{(1)}$ is the alpha heating of ordinary fuel and $R$ is the share of $P_\alpha^{(1)}$ lost to radiation. These are evaluated on the ignition boundary, not at the crossing, so they have no subscript $\times$. At fixed $T$ and $R$ the pressure needed to ignite therefore falls as $(1-R)/(A-R)$, and we define the pressure saving as
\begin{equation}
 \delta_{nT}\equiv1-\frac{(nT)(A)}{(nT)(1)}=\frac{A-1}{A-R}.
\label{eq:pressure}
\end{equation}
We call this the pressure rule. Polarization helps more where radiation consumes more of the alpha heating, a trend also seen for bremsstrahlung in Ref.~\cite{micklich1984} and for impurity and synchrotron radiation in Ref.~\cite{lfdav}.

\begin{table}[t]
\caption{Ignition access at $\tau_E=1$ s for the cases of Ref.~\cite{lfdav}: temperature $T_C$, heating $H_C$ and gain $Q_C$ at the Cordey pass of ordinary fuel, radiated share $R$ of the alpha power at the entry point, and the heating saving $\delta_H$ at the Cordey pass and the pressure saving $\delta_{nT}$ at the entry point, each at $A=1.2$ and 1.5.}
\label{tab:cases}
\begin{ruledtabular}
\begin{tabular}{lcccccc}
Case & $T_C$ & $H_C$ & $Q_C$ & $R$ & $\delta_H$ & $\delta_{nT}$\\
 & (keV) & (kW/m$^3$) & & & (\%) & (\%)\\
\hline
Ideal & 13.5 & 333 & 5.0 & 0 & 17, 33 & 17, 33\\
Radiating & 12.3 & 425 & 5.2 & 0.14 & 17, 33 & 18, 35\\
W 10 ppm & 14.4 & 523 & 6.2 & 0.30 & 19, 37 & 20, 39\\
W 50 ppm & 16.5 & 626 & 7.2 & 0.39 & 21, 40 & 23, 43\\
W 100 ppm & 18.7 & 763 & 8.4 & 0.49 & 23, 43 & 28, 49\\
\end{tabular}
\end{ruledtabular}
\end{table}

\begin{figure*}[t]
\centering
\begin{subfigure}[t]{0.325\textwidth}
\includegraphics[width=\linewidth]{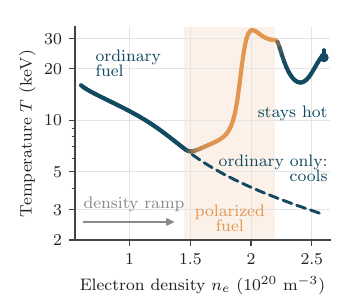}
\caption{Path to high gain at 400 kW/m$^3$}\label{fig:path-a}
\end{subfigure}\hfill
\begin{subfigure}[t]{0.325\textwidth}
\includegraphics[width=\linewidth]{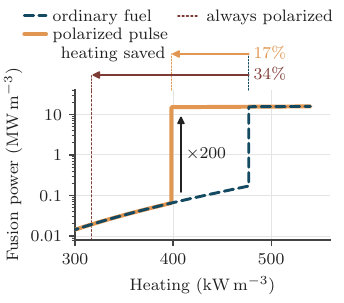}
\caption{}\label{fig:path-b}
\end{subfigure}\hfill
\begin{subfigure}[t]{0.325\textwidth}
\includegraphics[width=\linewidth]{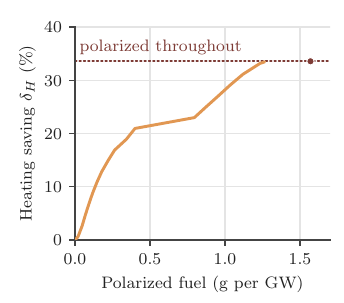}
\caption{}\label{fig:path-c}
\end{subfigure}
\caption{Reaching high gain with polarized fuel in the time-dependent model with $n_e$ ramped from 0.6 to $2.6\times10^{20}$ m$^{-3}$ in 75 s. (\subref{fig:path-a}) Temperature as the density increases, with ordinary fuel only (dashed) and with polarized fuel injected in the shaded interval. Circle: final state with ordinary fuel. (\subref{fig:path-b}) Fusion power after density ramp versus heating. Arrows show heating saved by the SPF pulse and by fuel that is always polarized (dotted). (\subref{fig:path-c}) Heating saving $\delta_H$ versus the fewest grams of polarized fuel per gigawatt of final fusion power that we found to bring plasma onto the hot branch with this density ramp.}
\label{fig:path}
\end{figure*}

We start with the simplest case of the `ideal' plasma, which has no ash and no radiation, because both of these rules are exact there. The auxiliary heating is $P_{\rm aux}=3nT/\tau_E-P_\alpha$ with $P_\alpha\propto An^2\langle\sigma v\rangle$, and setting both derivatives to zero gives $P_{\rm aux}=P_\alpha$ and $d\ln\langle\sigma v\rangle/d\ln T=2$, which D-T fuel satisfies at 13.5 keV~\cite{lfdav}. The gain at the Cordey pass is therefore $Q_C=5$ for all $A$ values. Here, the required heating at the Cordey pass decreases exactly as $1/A$, a decrease of a third for fully polarized fuel. Radiation increases the gain at the Cordey pass because the alpha heating must also replace the radiated power, and so this increases $\delta_H$, sometimes substantially: to 43\% with 100 ppm of tungsten (\cref{tab:cases}). \Cref{eq:rule} overestimates $dH_\times/dA$ by up to 13\%, because it neglects the effect of extra helium ash, and \cref{eq:saving} overestimates $\delta_H$ by 1 point in the radiating plasma and 7 points with 100 ppm of tungsten because the gain at the Cordey pass also decreases as $A$ increases. The entry pressure, the lowest $nT$ on the ignition boundary, falls by up to 49\%, within 3 points of \cref{eq:pressure}, and with 100 ppm of tungsten the pressure at the Cordey pass falls by a factor of 1.83, as in Ref.~\cite{lfdav}. 

The SPF benefits also persist with weaker polarization and different confinement times. Fuel with $A=1.2$ gives 40\% of the full enhancement, still saves about half as much as fully polarized fuel, and confinement times of 0.5 and 2 s change the heating at the Cordey pass of the radiating plasma by a factor of 12 but $\delta_H$ by at most a percentage point, a small effect. At the Cordey pass we find that polarization saves a third or more of the heating, and most where radiation makes high gain hardest to reach.

While these results are promising, there are also cases when SPF does not help much. For example, a plasma that accesses high gain by fueling to its operating density first and heating afterwards crosses at the cool-side maximum of $P_{\rm aux}(T)$, where the plasma gain is low. In our time-dependent model at $n_e=3\times10^{20}$ m$^{-3}$, ordinary fuel crosses with $Q_\times=2.1$ and fully polarized fuel lowers the heating needed by only 14\%. Fueling first also requires more heating: 686 kW/m$^3$ with ordinary fuel, 44\% more than the 477 kW/m$^3$ needed through the Cordey pass at lower density. Confinement that degrades with heating power, as in the ISS04 and IPB98(y,2) scalings~\cite{iss04,ipb98}, gives similar crossings at fixed density: at $n_e=1.6\times10^{20}$ m$^{-3}$ they have $Q_\times=2.8$ and 1.1, and polarization saves 17\% and 8\% [\cref{fig:access}(c)]. At the Cordey pass, power degradation has the opposite effect, because confinement improves as the heating falls. In the stationary model with $\tau_E\propto P^{-a}$ (total heating power $P$, $a$ from 0.3 to 0.67, $\tau_E=1$ s at 2 MW/m$^3$), $Q_C$ increases to between 11 and 24, and polarization removes 50 to 71\% of a heating at the Cordey pass that is already small, 7 to 210 kW/m$^3$. SPF therefore helps most through the Cordey pass and least in a plasma heated at fixed density.

In practice, two effects decrease $\delta_H$ at the Cordey pass below the values of \cref{tab:cases}. First, tungsten radiates most strongly near 1.5 to 2 keV~\cite{putterich} where alphas supply almost no heat. At the density of the Cordey pass with 100 ppm of tungsten, keeping the plasma at 1.7 keV requires 1550 kW/m$^3$, twice the heating at the Cordey pass, with or without polarization, so such a plasma must be heated through this peak at low density first. Second, fuel recycled from the wall returns depolarized~\cite{kvc,baylor} and dilutes the polarized core. To estimate the effect on $A$, we extended the TORAX transport code~\cite{torax} to include polarized fuel and simulated a tokamak with the size and field of ARC~\cite{sorbom2015} fueled by polarized pellets. The core reaches only $A=1.21$ to 1.32 for separatrix densities of 0.35 to 0.2 times the Greenwald density, which in the stationary calculation lowers $\delta_H$ at the Cordey pass of the radiating plasma from 33\% to between 17 and 25\%.

\section{Reaching high gain with polarized fuel}\label{sec:startup}

To find how many grams of polarized fuel a plasma needs to cross into high gain and when the polarized fuel should be supplied, we use the approach of Ref.~\cite{lfdav} in our time-dependent model: the plasma is heated at low density and then fueled up through the Cordey pass at constant heating. We increase $n_e$ from 0.6 to $2.6\times10^{20}$ m$^{-3}$ in 75 s, starting below the Cordey pass of polarized fuel, which is at a lower density than that of ordinary fuel by about the factor $A$. Ordinary fuel reaches high gain only above 477 kW/m$^3$ whereas fuel polarized throughout the density ramp needs 317 kW/m$^3$, 34\% less and close to the 33\% of \cref{tab:cases}.

The fuel needs to be polarized only while the plasma is near the Cordey pass (\cref{fig:path}). At 400 kW/m$^3$, a plasma on ordinary fuel cools to 2.8 keV as its density increases. If the fuel is polarized only while $n_e$ increases from 1.45 to $2.2\times10^{20}$ m$^{-3}$, for 28 s, the plasma heats up instead and stays hot after the fuel is ordinary again, at 23 keV and a gain of 38, with more than 200 times the fusion power [\cref{fig:path}(a) and (b)]. This plasma is not ignited, but its external heating is under 3\% of its fusion power and remains available to turn the fusion power down. 

\begin{figure*}[t]
\centering
\begin{subfigure}[t]{0.325\textwidth}
\includegraphics[width=\linewidth]{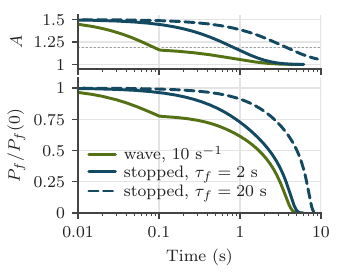}
\caption{Reactivity and fusion power}\label{fig:exit-a}
\end{subfigure}\hfill
\begin{subfigure}[t]{0.325\textwidth}
\includegraphics[width=\linewidth]{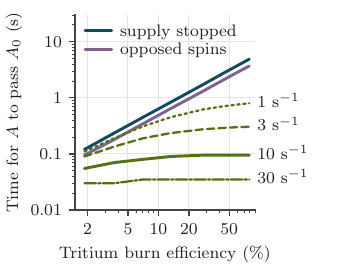}
\caption{Time to leave high gain}\label{fig:exit-b}
\end{subfigure}\hfill
\begin{subfigure}[t]{0.325\textwidth}
\includegraphics[width=\linewidth]{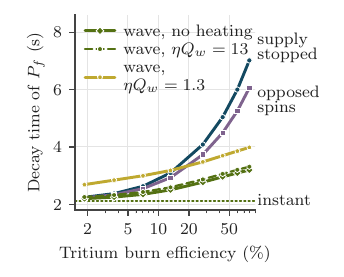}
\caption{Decay time of the fusion power}\label{fig:exit-c}
\end{subfigure}
\caption{Leaving ignition sustained by polarization, at $n_e=2.4\times10^{20}$ m$^{-3}$ without heating. (\subref{fig:exit-a}) $A$ and fusion power with forced depolarization at 10 s$^{-1}$ and after stopping SPF supply; dotted line: $A_0$. (\subref{fig:exit-b}) Time for $A$ to pass $A_0$ as a function of the tritium burn efficiency. (\subref{fig:exit-c}) Time for fusion power to fall by a factor of $e$, including the wave heating. Dotted line: an instantaneous, complete depolarization.}
\label{fig:exit}
\end{figure*}

This pulse uses 0.7 g of polarized fuel per gigawatt of fusion power in the final state, which produces a high power density of 15 MW/m$^3$ with a volume-averaged plasma $\beta$ up to 7.3\% at 10 T. This fuel consumption is less than one thousandth of the 2.7 kg per gigawatt that a day of continuous polarized fueling would use at this plasma's tritium burn efficiency of 10\%. The pulse lowers the heating needed from 477 to 398 kW/m$^3$, by 17\%. A pulse timed more closely to the crossing needs less, and the first tenths of a gram are the most important: 0.17 g per gigawatt lowers the heating by 12\% and 0.26 g by 17\%, whereas 0.40, 0.96 and 1.2 g, injected over a wider range of density, lower it by 21, 27 and 33\%, nearly the 34\% of fuel polarized throughout the density ramp [\cref{fig:path}(c)]. By fueling the plasma first and then heating (at fixed density), it needs even less polarized fuel, but much more heating: at $n_e=3\times10^{20}$ m$^{-3}$ and 637 kW/m$^3$, midway between the heating that ordinary and polarized fuel need, 7.4 s of polarized fueling, about 0.2 g per gigawatt, ignites it. The main condition of the density ramp is that it must be slow enough for the alpha heating to keep up, otherwise the added fuel cools the plasma faster than the alphas can heat it. With a density ramp twice as fast, the same pulse no longer brings the plasma to high gain.

\section{Leaving ignition}\label{sec:exit}

An ignited plasma is hard to turn down quickly, after a fault or in a thermal runaway, because the actuators that remain to it work only through its density and temperature, which change on slow confinement timescales. Here we propose that spin polarization is a different kind of actuator: it acts on the cross section itself, so removing it lowers the reactivity at once. This is clearest in a plasma that only polarization keeps ignited, between the ignition boundaries of ordinary and polarized fuel [\cref{fig:access}(a)], whose hot state exists only while $A$ stays above a value $A_0$ between 1 and 1.5. Such plasmas need continuous polarized fueling, the long-term aim for polarized fuel~\cite{parisi2024,baylor}. In our time-dependent model without auxiliary heating they exist for $n_e$ from about 1.9 to $2.9\times10^{20}$ m$^{-3}$, over which $A_0$ falls from 1.43 to 1.023. To leave ignition, the spin product $p_\mathrm{D}p_\mathrm{T}$ must decrease from 1 to $2(A_0-1)$, which is by 14\% at the lowest of these densities and by a factor of 22 at the highest.

A wave at the nuclear precession frequency $\omega=\gamma B$, where $\gamma$ is the gyromagnetic ratio of the nucleus and $B$ the field strength, depolarizes by magnetic resonance~\cite{abragam}. Because $A$ depends on the product $p_\mathrm{D}p_\mathrm{T}$, depolarizing one species is enough: tritons precess at 45.4 MHz/T and deuterons at 6.54 MHz/T, so at 10 T the deuteron frequency of 65 MHz is in the range of ion cyclotron heating systems, which use 40 to 55 MHz in ITER~\cite{lamalle2013} and 120 MHz in SPARC~\cite{lin2020}. A field of amplitude $b$ that rotates with the spins depolarizes them at the rate $\Gamma\simeq\pi\gamma b^2/\Delta B$, where $\Delta B$ is the variation of the field along the particle orbits~\cite{abragam}. We estimate this rate from the probability of a spin flip at each passage through resonance. This rate does not depend on particle speed because a faster particle crosses the resonance more often but is flipped less at each crossing. With $\Delta B=1$ T, $\Gamma=10$ s$^{-1}$ needs $b=0.11$ mT for tritons. In such a plasma at $n_e=2.4\times10^{20}$ m$^{-3}$, with $A_0=1.19$, a temperature of 26 keV and 18 MW/m$^3$ of fusion power, a wave at 10 s$^{-1}$ lowers the fusion power by a fifth and takes $A$ past $A_0$ within 0.1 s at every tritium burn efficiency [\cref{fig:exit}(a) and (b)]. From then on the plasma has no hot state, so exit is decided within a tenth of a second, although its fusion power takes 2.2 to 3.2 s to fall by a factor of $e$ [\cref{fig:exit}(c)]. A wave that stays on until the tritium is fully depolarized shortens the decay to 2.2 s at every tritium burn efficiency, close to the 2.1 s of an instantaneous depolarization. This is much slower than massive impurity injection~\cite{fpss}, so depolarization cannot replace an emergency shutdown. Instead it can steer the plasma, lowering the fusion power or ending high gain without a disruption, which a power plant must avoid~\cite{siccinio2020}.

Stopping the polarized supply lowers the reactivity only as fast as fuel exchange and fusion remove the polarized fuel: $A$ passes $A_0$ after 0.12 s at a tritium burn efficiency of 2\%, 0.83 s at 13\% and 4.9 s at 79\% [\cref{fig:exit}(b)], and the fusion power falls by a factor of $e$ after 2.2 to 7.0 s. The wave therefore helps most where the fuel is confined longest, at the high tritium burn efficiency that keeps the tritium inventory of a plant low~\cite{whyte2023}: at 79\% it shortens the decay from 7.0 to 3.2 s. These times are lower bounds, because a real fueling system takes time to switch.

The wave also heats the plasma. The energy stored in the wave is the absorbed power $P_w$ times the damping time $Q_w/\omega$, where the quality factor $Q_w$ is the number of radians of oscillation that the wave completes before it is absorbed. If a share $\eta$ of this energy is in the field that rotates with the spins, then $b^2=2\mu_0\eta Q_wP_w/(\omega V)$ for a plasma of volume $V$, with $\mu_0$ the vacuum permeability, and with $\omega=\gamma B$ the rate is $\Gamma=2\pi\mu_0\eta Q_wP_w/(B\,\Delta B\,V)$, so each reduction of $p_\mathrm{D}p_\mathrm{T}$ by a factor of $e$ costs the absorbed energy
\begin{equation}
 E_1=\frac{B\,\Delta B\,V}{2\pi\mu_0\eta Q_w},
\label{eq:cost}
\end{equation}
whatever the species or the rate: a fast depolarization needs a high power but no extra energy. At $B=10$ T and $\Delta B=1$ T, $\Gamma=10$ s$^{-1}$ requires $P_w\eta Q_w/V=13$ MW/m$^3$, for example $\eta Q_w=32$ with 0.4 MW/m$^3$ of wave power. Note the quantity $\eta Q_w$ is not a free parameter: once the wave mode is chosen, its field structure, its overlap with the orbits of the reacting ions, and its absorption fix $\eta Q_w$. We would need a full-wave code~\cite{brambilla1999,jaeger2001} to calculate $\eta Q_w$. We therefore treat $\eta Q_w$, and the rate $\Gamma$ throughout the plasma, as requirements on the wave. Rapidly depolarizing fuel across the whole plasma probably needs several frequencies or a frequency sweep.

This depolarization technique could halt a thermal runaway. At $n_e=2.4\times10^{20}$ m$^{-3}$ the polarized plasma also has a thermally unstable steady state, at 11 keV and 7.3 MW/m$^3$, and started 5\% above this temperature it runs away to 25 MW/m$^3$ within 5 s. If the polarized supply stops and a wave at 10 s$^{-1}$ starts when the fusion power reaches 9.5 MW/m$^3$, the fusion power increases no further and the plasma cools within 4 s for $\eta Q_w\geq13$. Feedback on the wave can also hold the plasma at an unstable state: with the depolarization rate proportional to the excess fusion power and $\eta Q_w=41$, the fusion power returns within 3 s of a 10\% increase in temperature at states between 12.5 and 17 keV. This feedback needs an efficient wave: it works for $\eta Q_w$ of about 30 or more and fails at 13, where the wave's heating outweighs the reactivity it removes.

\begin{figure}[t]
\centering
\includegraphics[width=\columnwidth]{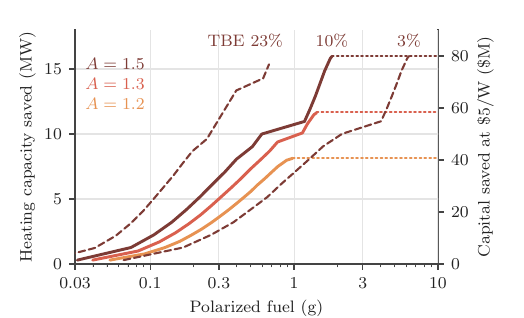}
\caption{Heating capacity saved by polarized pellets in a 100 m$^3$ plasma producing 1.5 GW of fusion power, and its initial capital value at \$5 per watt (2010 dollars)~\cite{sheffield2016,wade2021}, as a function of the polarized fuel used, in the density-ramp simulations of \cref{fig:path}. $A=1.5$ is a fully polarized core, and $A=1.3$ and 1.2 are cores diluted by depolarized fuel from the edge, all at a tritium burn efficiency (TBE) of 10\%. Dashed lines are $A=1.5$ at a TBE of 23\% and of 3\%. Dotted lines mark the most that each can save.}
\label{fig:value}
\end{figure}

A plasma that ordinary fuel keeps ignited has no polarization to remove, so a wave cannot take it out of ignition, but fueling with opposed spins can. At $n_e=3\times10^{20}$ m$^{-3}$, where ordinary fuel is ignited at 21 keV, 1.4 s of fueling with deuteron spins along the field and triton spins opposite to it takes the plasma out of ignition, and the spin direction of each isotope can be chosen when the fuel is polarized~\cite{ciullo2016,spiliotis2021}. A plant could therefore enter ignition with a pulse of aligned fuel and leave it with a pulse of opposed fuel.

\section{Discussion}\label{sec:summary}

We have shown that even grams of spin-polarized fuel (SPF) could be extremely valuable for tokamak and stellarator operation: a small pulse of SPF can grant access to a high-gain plasma, and stay there, even as the SPF is replaced by regular unpolarized fuel. This quantifies the transient use of polarized fuel during startup proposed by Mitarai, Hasuyama, and Wakuta~\cite{mitarai}. In practice, this can significantly reduce the required heating installed on a machine (\cref{fig:value}). This heating reduction is most applicable when high-gain access sets the installed heating requirement. In a tokamak, the installed heating may instead be set by current drive for steady-state operation, by the power needed to reach H-mode, or by a margin for control or design uncertainty, so stellarators, which need no current drive, probably benefit more. Additionally, because a burning tokamak or stellarator plasma has yet to be operated experimentally, there is significant scientific uncertainty for high-gain access. SPF partially derisks this uncertainty by increasing the probability the machine can access the high-gain regimes for high fusion power and net electricity. Another benefit includes less first-wall space occupied by ports for heating systems, increasing the tritium breeding ratio and simplifying design.

Therefore, a program to develop SPF does not need to produce kilograms of SPF per day per power plant to be useful. SPF can be useful even in \textit{grams}. For continuous-operation machines that operate for weeks, months, or even years without downtime, this would require many orders of magnitude less SPF than for continuous SPF operation. Even for pulsed machines such as ARC V3A, with a 900 s flat top~\cite{hillesheim2026overview}, a pulse would use tens of times less SPF than polarizing the fuel throughout the flat top: the ratio is roughly that of the flat-top duration to the pulse duration, whatever the tritium burn efficiency, because both amounts scale inversely with it. Finally, before power plants are deployed, small quantities of SPF can also de-risk access to burning plasma physics experiments by increasing the probability that high-gain is accessible on these devices. Accessing and controlling burning plasmas is crucial as we scale to power plants.

Polarization may also make high-gain and ignition easier to control. In a plasma that only polarization keeps ignited, forced depolarization lowers the reactivity at once and allows a faster exiting of the high-gain regime. It is slower than massive impurity injection, so it is a way of steering the plasma rather than emergency shutdown. It may also stop a thermal runaway, and under feedback it can hold a plasma at a thermally unstable state. The wave adds little heat to a high-gain exit if it couples efficiently to the nuclear spins, and fueling with anti-aligned nuclear spins takes a plasma that ordinary fuel keeps ignited out of ignition. 

These benefits could also extend to other machines such as tandem mirrors, where polarized fuel has been studied before~\cite{dinicola1987,noever1995}. We calculated the zero-dimensional power balance of the axisymmetric tandem mirror of Frank \textit{et al.}~\cite{frank2025,frank2026}, whose end plugs confine the ions~\cite{post1987}. We found a polarized pulse does not help much because the central cell needs the most heating while it is cold, and a depolarizing wave is not needed for control, because the plug potential turns the central cell down faster. Continuous polarization does still help: at $A=1.5$ and the density and plasma pressure of the design point, it increases the fusion power by 56\% and the gain (counting the plug neutral beams) by a factor of 1.8.

There are, of course, significant uncertainties. First is that SPF does not work as expected. This could include it depolarizing too quickly or being hard to make at high polarization fraction and/or rate. Second is that all of the predicted benefits conferred by SPF are simply not useful for fusion deployment, and that the most challenging aspects of deployment are limited by things not helped by SPF. However, given the wide range of benefits from increased tritium burn efficiency, to high power density, to anisotropic neutron emission, to relaxed confinement requirements, it is hard to see SPF not being useful.

\begin{acknowledgments}
I thank Adam Rutkowski for helpful conversations.
\end{acknowledgments}

\section*{Data availability}

The data that support the findings of this study, and the scripts that produce them, will be made publicly available upon publication.

\section*{AI statement}

I made extensive use of Anthropic's Claude Opus 5.5 model in all aspects of this work. I take full responsibility for its content.

\bibliography{spf_dose_nf_letter}

\begin{thebibliography}{75}%
\makeatletter
\providecommand \@ifxundefined [1]{%
 \@ifx{#1\undefined}
}%
\providecommand \@ifnum [1]{%
 \ifnum #1\expandafter \@firstoftwo
 \else \expandafter \@secondoftwo
 \fi
}%
\providecommand \@ifx [1]{%
 \ifx #1\expandafter \@firstoftwo
 \else \expandafter \@secondoftwo
 \fi
}%
\providecommand \natexlab [1]{#1}%
\providecommand \enquote  [1]{``#1''}%
\providecommand \bibnamefont  [1]{#1}%
\providecommand \bibfnamefont [1]{#1}%
\providecommand \citenamefont [1]{#1}%
\providecommand \href@noop [0]{\@secondoftwo}%
\providecommand \href [0]{\begingroup \@sanitize@url \@href}%
\providecommand \@href[1]{\@@startlink{#1}\@@href}%
\providecommand \@@href[1]{\endgroup#1\@@endlink}%
\providecommand \@sanitize@url [0]{\catcode `\\12\catcode `\$12\catcode
  `\&12\catcode `\#12\catcode `\^12\catcode `\_12\catcode `\%12\relax}%
\providecommand \@@startlink[1]{}%
\providecommand \@@endlink[0]{}%
\providecommand \url  [0]{\begingroup\@sanitize@url \@url }%
\providecommand \@url [1]{\endgroup\@href {#1}{\urlprefix }}%
\providecommand \urlprefix  [0]{URL }%
\providecommand \Eprint [0]{\href }%
\providecommand \doibase [0]{https://doi.org/}%
\providecommand \selectlanguage [0]{\@gobble}%
\providecommand \bibinfo  [0]{\@secondoftwo}%
\providecommand \bibfield  [0]{\@secondoftwo}%
\providecommand \translation [1]{[#1]}%
\providecommand \BibitemOpen [0]{}%
\providecommand \bibitemStop [0]{}%
\providecommand \bibitemNoStop [0]{.\EOS\space}%
\providecommand \EOS [0]{\spacefactor3000\relax}%
\providecommand \BibitemShut  [1]{\csname bibitem#1\endcsname}%
\let\auto@bib@innerbib\@empty
\bibitem [{\citenamefont {Lawson}(1957)}]{lawson}%
  \BibitemOpen
  \bibfield  {author} {\bibinfo {author} {\bibfnamefont {J.~D.}\ \bibnamefont
  {Lawson}},\ }\bibfield  {title} {\bibinfo {title} {Some criteria for a power
  producing thermonuclear reactor},\ }\href
  {https://doi.org/10.1088/0370-1301/70/1/303} {\bibfield  {journal} {\bibinfo
  {journal} {Proc. Phys. Soc. B}\ }\textbf {\bibinfo {volume} {70}},\ \bibinfo
  {pages} {6} (\bibinfo {year} {1957})}\BibitemShut {NoStop}%
\bibitem [{\citenamefont {Wurzel}\ and\ \citenamefont
  {Hsu}(2022)}]{wurzel2022}%
  \BibitemOpen
  \bibfield  {author} {\bibinfo {author} {\bibfnamefont {S.~E.}\ \bibnamefont
  {Wurzel}}\ and\ \bibinfo {author} {\bibfnamefont {S.~C.}\ \bibnamefont
  {Hsu}},\ }\bibfield  {title} {\bibinfo {title} {Progress toward fusion energy
  breakeven and gain as measured against the {L}awson criterion},\ }\href
  {https://doi.org/10.1063/5.0083990} {\bibfield  {journal} {\bibinfo
  {journal} {Phys. Plasmas}\ }\textbf {\bibinfo {volume} {29}},\ \bibinfo
  {pages} {062103} (\bibinfo {year} {2022})}\BibitemShut {NoStop}%
\bibitem [{\citenamefont {Shimada}\ \emph {et~al.}(2007)\citenamefont
  {Shimada}, \citenamefont {Campbell}, \citenamefont {Mukhovatov},
  \citenamefont {Fujiwara}, \citenamefont {Kirneva} \emph
  {et~al.}}]{shimada2007}%
  \BibitemOpen
  \bibfield  {author} {\bibinfo {author} {\bibfnamefont {M.}~\bibnamefont
  {Shimada}}, \bibinfo {author} {\bibfnamefont {D.~J.}\ \bibnamefont
  {Campbell}}, \bibinfo {author} {\bibfnamefont {V.}~\bibnamefont
  {Mukhovatov}}, \bibinfo {author} {\bibfnamefont {M.}~\bibnamefont
  {Fujiwara}}, \bibinfo {author} {\bibfnamefont {N.}~\bibnamefont {Kirneva}},
  \emph {et~al.},\ }\bibfield  {title} {\bibinfo {title} {Progress in the
  {ITER} physics basis {Chapter} 1: Overview and summary},\ }\href
  {https://doi.org/10.1088/0029-5515/47/6/S01} {\bibfield  {journal} {\bibinfo
  {journal} {Nucl. Fusion}\ }\textbf {\bibinfo {volume} {47}},\ \bibinfo
  {pages} {S1} (\bibinfo {year} {2007})}\BibitemShut {NoStop}%
\bibitem [{\citenamefont {Creely}\ \emph {et~al.}(2020)\citenamefont {Creely},
  \citenamefont {Greenwald}, \citenamefont {Ballinger}, \citenamefont
  {Brunner}, \citenamefont {Canik} \emph {et~al.}}]{creely2020}%
  \BibitemOpen
  \bibfield  {author} {\bibinfo {author} {\bibfnamefont {A.~J.}\ \bibnamefont
  {Creely}}, \bibinfo {author} {\bibfnamefont {M.~J.}\ \bibnamefont
  {Greenwald}}, \bibinfo {author} {\bibfnamefont {S.~B.}\ \bibnamefont
  {Ballinger}}, \bibinfo {author} {\bibfnamefont {D.}~\bibnamefont {Brunner}},
  \bibinfo {author} {\bibfnamefont {J.}~\bibnamefont {Canik}}, \emph {et~al.},\
  }\bibfield  {title} {\bibinfo {title} {Overview of the {SPARC} tokamak},\
  }\href {https://doi.org/10.1017/S0022377820001257} {\bibfield  {journal}
  {\bibinfo  {journal} {J. Plasma Phys.}\ }\textbf {\bibinfo {volume} {86}},\
  \bibinfo {pages} {865860502} (\bibinfo {year} {2020})}\BibitemShut {NoStop}%
\bibitem [{\citenamefont {Hillesheim}\ \emph {et~al.}(2026)\citenamefont
  {Hillesheim}, \citenamefont {Creely}, \citenamefont {Eich}, \citenamefont
  {Howard}, \citenamefont {Leuthold}, \citenamefont {Sweeney}, \citenamefont
  {LeViness}, \citenamefont {Nelson}, \citenamefont {Nichols}, \citenamefont
  {Tinguely} \emph {et~al.}}]{hillesheim2026overview}%
  \BibitemOpen
  \bibfield  {author} {\bibinfo {author} {\bibfnamefont {J.~C.}\ \bibnamefont
  {Hillesheim}}, \bibinfo {author} {\bibfnamefont {A.~J.}\ \bibnamefont
  {Creely}}, \bibinfo {author} {\bibfnamefont {T.~H.}\ \bibnamefont {Eich}},
  \bibinfo {author} {\bibfnamefont {N.~T.}\ \bibnamefont {Howard}}, \bibinfo
  {author} {\bibfnamefont {N.}~\bibnamefont {Leuthold}}, \bibinfo {author}
  {\bibfnamefont {R.}~\bibnamefont {Sweeney}}, \bibinfo {author} {\bibfnamefont
  {A.}~\bibnamefont {LeViness}}, \bibinfo {author} {\bibfnamefont {A.~O.}\
  \bibnamefont {Nelson}}, \bibinfo {author} {\bibfnamefont {L.}~\bibnamefont
  {Nichols}}, \bibinfo {author} {\bibfnamefont {R.~A.}\ \bibnamefont
  {Tinguely}}, \emph {et~al.},\ }\bibfield  {title} {\bibinfo {title} {Overview
  of the physics basis for the arc fusion power plant},\ }\href
  {https://doi.org/10.1017/S0022377826101706} {\bibfield  {journal} {\bibinfo
  {journal} {Journal of Plasma Physics}\ }\textbf {\bibinfo {volume} {92}},\
  \bibinfo {pages} {E69} (\bibinfo {year} {2026})}\BibitemShut {NoStop}%
\bibitem [{\citenamefont {Meyer}\ and\ \citenamefont {{STEP Plasma
  Team}}(2024)}]{meyer2024}%
  \BibitemOpen
  \bibfield  {author} {\bibinfo {author} {\bibfnamefont {H.}~\bibnamefont
  {Meyer}}\ and\ \bibinfo {author} {\bibnamefont {{STEP Plasma Team}}},\
  }\bibfield  {title} {\bibinfo {title} {Plasma burn{\textemdash}mind the
  gap},\ }\href {https://doi.org/10.1098/rsta.2023.0406} {\bibfield  {journal}
  {\bibinfo  {journal} {Phil. Trans. R. Soc. A}\ }\textbf {\bibinfo {volume}
  {382}},\ \bibinfo {pages} {20230406} (\bibinfo {year} {2024})}\BibitemShut
  {NoStop}%
\bibitem [{\citenamefont {Willis}\ \emph {et~al.}(2026)\citenamefont {Willis},
  \citenamefont {McNamara}, \citenamefont {Maartensson}, \citenamefont
  {Astbury}, \citenamefont {Yildirim} \emph {et~al.}}]{willis2026}%
  \BibitemOpen
  \bibfield  {author} {\bibinfo {author} {\bibfnamefont {J.}~\bibnamefont
  {Willis}}, \bibinfo {author} {\bibfnamefont {S.~A.~M.}\ \bibnamefont
  {McNamara}}, \bibinfo {author} {\bibfnamefont {E.~N.~J.}\ \bibnamefont
  {Maartensson}}, \bibinfo {author} {\bibfnamefont {J.}~\bibnamefont
  {Astbury}}, \bibinfo {author} {\bibfnamefont {E.}~\bibnamefont {Yildirim}},
  \emph {et~al.},\ }\bibfield  {title} {\bibinfo {title} {Tokamak {Energy}'s
  pre-concept design for a fusion power plant: an overview of {ST-E1}},\ }\href
  {https://doi.org/10.1088/1741-4326/ae5545} {\bibfield  {journal} {\bibinfo
  {journal} {Nucl. Fusion}\ }\textbf {\bibinfo {volume} {66}},\ \bibinfo
  {pages} {086001} (\bibinfo {year} {2026})}\BibitemShut {NoStop}%
\bibitem [{\citenamefont {{The MANTA Collaboration}}\ \emph
  {et~al.}(2024)\citenamefont {{The MANTA Collaboration}}, \citenamefont
  {Rutherford}, \citenamefont {Wilson}, \citenamefont {Saltzman}, \citenamefont
  {Arnold} \emph {et~al.}}]{manta2024}%
  \BibitemOpen
  \bibfield  {author} {\bibinfo {author} {\bibnamefont {{The MANTA
  Collaboration}}}, \bibinfo {author} {\bibfnamefont {G.}~\bibnamefont
  {Rutherford}}, \bibinfo {author} {\bibfnamefont {H.~S.}\ \bibnamefont
  {Wilson}}, \bibinfo {author} {\bibfnamefont {A.}~\bibnamefont {Saltzman}},
  \bibinfo {author} {\bibfnamefont {D.}~\bibnamefont {Arnold}}, \emph
  {et~al.},\ }\bibfield  {title} {\bibinfo {title} {{MANTA}: a
  negative-triangularity {NASEM}-compliant fusion pilot plant},\ }\href
  {https://doi.org/10.1088/1361-6587/ad6708} {\bibfield  {journal} {\bibinfo
  {journal} {Plasma Phys. Control. Fusion}\ }\textbf {\bibinfo {volume} {66}},\
  \bibinfo {pages} {105006} (\bibinfo {year} {2024})}\BibitemShut {NoStop}%
\bibitem [{\citenamefont {Brown}\ \emph {et~al.}(2026)\citenamefont {Brown},
  \citenamefont {Menard}, \citenamefont {Berkery}, \citenamefont {Brooks},
  \citenamefont {Costley} \emph {et~al.}}]{brown2026}%
  \BibitemOpen
  \bibfield  {author} {\bibinfo {author} {\bibfnamefont {T.~G.}\ \bibnamefont
  {Brown}}, \bibinfo {author} {\bibfnamefont {J.~E.}\ \bibnamefont {Menard}},
  \bibinfo {author} {\bibfnamefont {J.~W.}\ \bibnamefont {Berkery}}, \bibinfo
  {author} {\bibfnamefont {A.}~\bibnamefont {Brooks}}, \bibinfo {author}
  {\bibfnamefont {A.~E.}\ \bibnamefont {Costley}}, \emph {et~al.},\ }\bibfield
  {title} {\bibinfo {title} {The spherical tokamak advanced reactor ({STAR})
  fusion power plant design},\ }\href
  {https://doi.org/10.1016/j.fusengdes.2025.115461} {\bibfield  {journal}
  {\bibinfo  {journal} {Fusion Eng. Des.}\ }\textbf {\bibinfo {volume} {222}},\
  \bibinfo {pages} {115461} (\bibinfo {year} {2026})}\BibitemShut {NoStop}%
\bibitem [{\citenamefont {Siccinio}\ \emph {et~al.}(2020)\citenamefont
  {Siccinio}, \citenamefont {Biel}, \citenamefont {Cavedon}, \citenamefont
  {Fable}, \citenamefont {Federici}, \citenamefont {Janky} \emph
  {et~al.}}]{siccinio2020}%
  \BibitemOpen
  \bibfield  {author} {\bibinfo {author} {\bibfnamefont {M.}~\bibnamefont
  {Siccinio}}, \bibinfo {author} {\bibfnamefont {W.}~\bibnamefont {Biel}},
  \bibinfo {author} {\bibfnamefont {M.}~\bibnamefont {Cavedon}}, \bibinfo
  {author} {\bibfnamefont {E.}~\bibnamefont {Fable}}, \bibinfo {author}
  {\bibfnamefont {G.}~\bibnamefont {Federici}}, \bibinfo {author}
  {\bibfnamefont {F.}~\bibnamefont {Janky}}, \emph {et~al.},\ }\bibfield
  {title} {\bibinfo {title} {{DEMO} physics challenges beyond {ITER}},\ }\href
  {https://doi.org/10.1016/j.fusengdes.2020.111603} {\bibfield  {journal}
  {\bibinfo  {journal} {Fusion Eng. Des.}\ }\textbf {\bibinfo {volume} {156}},\
  \bibinfo {pages} {111603} (\bibinfo {year} {2020})}\BibitemShut {NoStop}%
\bibitem [{\citenamefont {Zhuang}\ \emph {et~al.}(2019)\citenamefont {Zhuang},
  \citenamefont {Li}, \citenamefont {Li}, \citenamefont {Wan}, \citenamefont
  {Liu} \emph {et~al.}}]{zhuang2019}%
  \BibitemOpen
  \bibfield  {author} {\bibinfo {author} {\bibfnamefont {G.}~\bibnamefont
  {Zhuang}}, \bibinfo {author} {\bibfnamefont {G.~Q.}\ \bibnamefont {Li}},
  \bibinfo {author} {\bibfnamefont {J.}~\bibnamefont {Li}}, \bibinfo {author}
  {\bibfnamefont {Y.~X.}\ \bibnamefont {Wan}}, \bibinfo {author} {\bibfnamefont
  {Y.}~\bibnamefont {Liu}}, \emph {et~al.},\ }\bibfield  {title} {\bibinfo
  {title} {Progress of the {CFETR} design},\ }\href
  {https://doi.org/10.1088/1741-4326/ab0e27} {\bibfield  {journal} {\bibinfo
  {journal} {Nucl. Fusion}\ }\textbf {\bibinfo {volume} {59}},\ \bibinfo
  {pages} {112010} (\bibinfo {year} {2019})}\BibitemShut {NoStop}%
\bibitem [{\citenamefont {Kim}\ \emph {et~al.}(2015)\citenamefont {Kim},
  \citenamefont {Im}, \citenamefont {Kim}, \citenamefont {Oh}, \citenamefont
  {Park} \emph {et~al.}}]{kim2015}%
  \BibitemOpen
  \bibfield  {author} {\bibinfo {author} {\bibfnamefont {K.}~\bibnamefont
  {Kim}}, \bibinfo {author} {\bibfnamefont {K.}~\bibnamefont {Im}}, \bibinfo
  {author} {\bibfnamefont {H.~C.}\ \bibnamefont {Kim}}, \bibinfo {author}
  {\bibfnamefont {S.}~\bibnamefont {Oh}}, \bibinfo {author} {\bibfnamefont
  {J.~S.}\ \bibnamefont {Park}}, \emph {et~al.},\ }\bibfield  {title} {\bibinfo
  {title} {Design concept of {K-DEMO} for near-term implementation},\ }\href
  {https://doi.org/10.1088/0029-5515/55/5/053027} {\bibfield  {journal}
  {\bibinfo  {journal} {Nucl. Fusion}\ }\textbf {\bibinfo {volume} {55}},\
  \bibinfo {pages} {053027} (\bibinfo {year} {2015})}\BibitemShut {NoStop}%
\bibitem [{\citenamefont {Tobita}\ \emph {et~al.}(2019)\citenamefont {Tobita},
  \citenamefont {Hiwatari}, \citenamefont {Sakamoto}, \citenamefont {Someya},
  \citenamefont {Asakura} \emph {et~al.}}]{tobita2019}%
  \BibitemOpen
  \bibfield  {author} {\bibinfo {author} {\bibfnamefont {K.}~\bibnamefont
  {Tobita}}, \bibinfo {author} {\bibfnamefont {R.}~\bibnamefont {Hiwatari}},
  \bibinfo {author} {\bibfnamefont {Y.}~\bibnamefont {Sakamoto}}, \bibinfo
  {author} {\bibfnamefont {Y.}~\bibnamefont {Someya}}, \bibinfo {author}
  {\bibfnamefont {N.}~\bibnamefont {Asakura}}, \emph {et~al.},\ }\bibfield
  {title} {\bibinfo {title} {Japan's efforts to develop the concept of {JA
  DEMO} during the past decade},\ }\href
  {https://doi.org/10.1080/15361055.2019.1600931} {\bibfield  {journal}
  {\bibinfo  {journal} {Fusion Sci. Technol.}\ }\textbf {\bibinfo {volume}
  {75}},\ \bibinfo {pages} {372} (\bibinfo {year} {2019})}\BibitemShut
  {NoStop}%
\bibitem [{\citenamefont {Lion}\ \emph {et~al.}(2025)\citenamefont {Lion},
  \citenamefont {Angl{\`e}s}, \citenamefont {Bonauer}, \citenamefont
  {Ba{\~n}{\'o}n~Navarro}, \citenamefont {Cadena~Ceron} \emph
  {et~al.}}]{lion2025}%
  \BibitemOpen
  \bibfield  {author} {\bibinfo {author} {\bibfnamefont {J.}~\bibnamefont
  {Lion}}, \bibinfo {author} {\bibfnamefont {J.-C.}\ \bibnamefont
  {Angl{\`e}s}}, \bibinfo {author} {\bibfnamefont {L.}~\bibnamefont {Bonauer}},
  \bibinfo {author} {\bibfnamefont {A.}~\bibnamefont {Ba{\~n}{\'o}n~Navarro}},
  \bibinfo {author} {\bibfnamefont {S.~A.}\ \bibnamefont {Cadena~Ceron}}, \emph
  {et~al.},\ }\bibfield  {title} {\bibinfo {title} {Stellaris: A high-field
  quasi-isodynamic stellarator for a prototypical fusion power plant},\ }\href
  {https://doi.org/10.1016/j.fusengdes.2025.114868} {\bibfield  {journal}
  {\bibinfo  {journal} {Fusion Eng. Des.}\ }\textbf {\bibinfo {volume} {214}},\
  \bibinfo {pages} {114868} (\bibinfo {year} {2025})}\BibitemShut {NoStop}%
\bibitem [{\citenamefont {Hegna}\ \emph {et~al.}(2025)\citenamefont {Hegna},
  \citenamefont {Anderson}, \citenamefont {Andrew}, \citenamefont {Ayilaran},
  \citenamefont {Bader} \emph {et~al.}}]{hegna2025}%
  \BibitemOpen
  \bibfield  {author} {\bibinfo {author} {\bibfnamefont {C.~C.}\ \bibnamefont
  {Hegna}}, \bibinfo {author} {\bibfnamefont {D.~T.}\ \bibnamefont {Anderson}},
  \bibinfo {author} {\bibfnamefont {E.~C.}\ \bibnamefont {Andrew}}, \bibinfo
  {author} {\bibfnamefont {A.}~\bibnamefont {Ayilaran}}, \bibinfo {author}
  {\bibfnamefont {A.}~\bibnamefont {Bader}}, \emph {et~al.},\ }\bibfield
  {title} {\bibinfo {title} {The {Infinity Two} fusion pilot plant baseline
  plasma physics design},\ }\href {https://doi.org/10.1017/S0022377825000364}
  {\bibfield  {journal} {\bibinfo  {journal} {J. Plasma Phys.}\ }\textbf
  {\bibinfo {volume} {91}},\ \bibinfo {pages} {E76} (\bibinfo {year}
  {2025})}\BibitemShut {NoStop}%
\bibitem [{\citenamefont {Swanson}\ \emph {et~al.}(2026)\citenamefont
  {Swanson}, \citenamefont {Kumar}, \citenamefont {Dudt}, \citenamefont {Flom},
  \citenamefont {Kalb} \emph {et~al.}}]{swanson2026}%
  \BibitemOpen
  \bibfield  {author} {\bibinfo {author} {\bibfnamefont {C.~P.~S.}\
  \bibnamefont {Swanson}}, \bibinfo {author} {\bibfnamefont {S.~T.~A.}\
  \bibnamefont {Kumar}}, \bibinfo {author} {\bibfnamefont {D.~W.}\ \bibnamefont
  {Dudt}}, \bibinfo {author} {\bibfnamefont {E.~R.}\ \bibnamefont {Flom}},
  \bibinfo {author} {\bibfnamefont {W.~B.}\ \bibnamefont {Kalb}}, \emph
  {et~al.},\ }\bibfield  {title} {\bibinfo {title} {Overview of the {Helios}
  design: A practical planar coil stellarator fusion power plant},\ }\href
  {https://doi.org/10.1016/j.fusengdes.2026.115999} {\bibfield  {journal}
  {\bibinfo  {journal} {Fusion Eng. Des.}\ }\textbf {\bibinfo {volume} {232}},\
  \bibinfo {pages} {115999} (\bibinfo {year} {2026})}\BibitemShut {NoStop}%
\bibitem [{\citenamefont {Lazerson}\ \emph {et~al.}(2026)\citenamefont
  {Lazerson}, \citenamefont {Coelho}, \citenamefont {Douqa}, \citenamefont
  {Fessler}, \citenamefont {H{\"u}bner} \emph {et~al.}}]{lazerson2026}%
  \BibitemOpen
  \bibfield  {author} {\bibinfo {author} {\bibfnamefont {S.~A.}\ \bibnamefont
  {Lazerson}}, \bibinfo {author} {\bibfnamefont {A.~J.}\ \bibnamefont
  {Coelho}}, \bibinfo {author} {\bibfnamefont {D.}~\bibnamefont {Douqa}},
  \bibinfo {author} {\bibfnamefont {A.~A.}\ \bibnamefont {Fessler}}, \bibinfo
  {author} {\bibfnamefont {L.}~\bibnamefont {H{\"u}bner}}, \emph {et~al.},\
  }\href@noop {} {\bibinfo {title} {The fixed boundary plasma equilibrium basis
  for a one {Gigawatt} electric stellarator power plant}} (\bibinfo {year}
  {2026}),\ \Eprint {https://arxiv.org/abs/2607.09346} {arXiv:2607.09346}
  \BibitemShut {NoStop}%
\bibitem [{\citenamefont {Prost}\ and\ \citenamefont
  {Volpe}(2024)}]{prost2024}%
  \BibitemOpen
  \bibfield  {author} {\bibinfo {author} {\bibfnamefont {V.}~\bibnamefont
  {Prost}}\ and\ \bibinfo {author} {\bibfnamefont {F.~A.}\ \bibnamefont
  {Volpe}},\ }\bibfield  {title} {\bibinfo {title} {Economically optimized
  design point of high-field stellarator power-plant},\ }\href
  {https://doi.org/10.1088/1741-4326/ad142e} {\bibfield  {journal} {\bibinfo
  {journal} {Nucl. Fusion}\ }\textbf {\bibinfo {volume} {64}},\ \bibinfo
  {pages} {026007} (\bibinfo {year} {2024})}\BibitemShut {NoStop}%
\bibitem [{\citenamefont {Muldrew}\ \emph {et~al.}(2021)\citenamefont {Muldrew}
  \emph {et~al.}}]{helias}%
  \BibitemOpen
  \bibfield  {author} {\bibinfo {author} {\bibfnamefont {S.~I.}\ \bibnamefont
  {Muldrew}} \emph {et~al.},\ }\bibfield  {title} {\bibinfo {title} {Design
  uncertainty for a {HELIAS} 5-{B} stellarator fusion power plant},\ }\href
  {https://doi.org/10.1016/j.fusengdes.2021.112708} {\bibfield  {journal}
  {\bibinfo  {journal} {Fusion Eng. Des.}\ }\textbf {\bibinfo {volume} {170}},\
  \bibinfo {pages} {112708} (\bibinfo {year} {2021})}\BibitemShut {NoStop}%
\bibitem [{\citenamefont {Sagara}\ \emph {et~al.}(2014)\citenamefont {Sagara},
  \citenamefont {Tamura}, \citenamefont {Tanaka}, \citenamefont {Yanagi},
  \citenamefont {Miyazawa} \emph {et~al.}}]{sagara2014}%
  \BibitemOpen
  \bibfield  {author} {\bibinfo {author} {\bibfnamefont {A.}~\bibnamefont
  {Sagara}}, \bibinfo {author} {\bibfnamefont {H.}~\bibnamefont {Tamura}},
  \bibinfo {author} {\bibfnamefont {T.}~\bibnamefont {Tanaka}}, \bibinfo
  {author} {\bibfnamefont {N.}~\bibnamefont {Yanagi}}, \bibinfo {author}
  {\bibfnamefont {J.}~\bibnamefont {Miyazawa}}, \emph {et~al.},\ }\bibfield
  {title} {\bibinfo {title} {Helical reactor design {FFHR-d1} and c1 for
  steady-state {DEMO}},\ }\href
  {https://doi.org/10.1016/j.fusengdes.2014.02.076} {\bibfield  {journal}
  {\bibinfo  {journal} {Fusion Eng. Des.}\ }\textbf {\bibinfo {volume} {89}},\
  \bibinfo {pages} {2114} (\bibinfo {year} {2014})}\BibitemShut {NoStop}%
\bibitem [{\citenamefont {{ITER Physics Expert Group on Confinement and
  Transport}}\ \emph {et~al.}(1999)\citenamefont {{ITER Physics Expert Group on
  Confinement and Transport}}, \citenamefont {{ITER Physics Expert Group on
  Confinement Modelling and Database}},\ and\ \citenamefont {{ITER Physics
  Basis Editors}}}]{ipb98}%
  \BibitemOpen
  \bibfield  {author} {\bibinfo {author} {\bibnamefont {{ITER Physics Expert
  Group on Confinement and Transport}}}, \bibinfo {author} {\bibnamefont {{ITER
  Physics Expert Group on Confinement Modelling and Database}}},\ and\ \bibinfo
  {author} {\bibnamefont {{ITER Physics Basis Editors}}},\ }\bibfield  {title}
  {\bibinfo {title} {Chapter 2: Plasma confinement and transport},\ }\href
  {https://doi.org/10.1088/0029-5515/39/12/302} {\bibfield  {journal} {\bibinfo
   {journal} {Nucl. Fusion}\ }\textbf {\bibinfo {volume} {39}},\ \bibinfo
  {pages} {2175} (\bibinfo {year} {1999})}\BibitemShut {NoStop}%
\bibitem [{\citenamefont {Verdoolaege}\ \emph {et~al.}(2021)\citenamefont
  {Verdoolaege}, \citenamefont {Kaye}, \citenamefont {Angioni}, \citenamefont
  {Kardaun}, \citenamefont {Maslov}, \citenamefont {Romanelli} \emph
  {et~al.}}]{verdoolaege2021}%
  \BibitemOpen
  \bibfield  {author} {\bibinfo {author} {\bibfnamefont {G.}~\bibnamefont
  {Verdoolaege}}, \bibinfo {author} {\bibfnamefont {S.~M.}\ \bibnamefont
  {Kaye}}, \bibinfo {author} {\bibfnamefont {C.}~\bibnamefont {Angioni}},
  \bibinfo {author} {\bibfnamefont {O.~J. W.~F.}\ \bibnamefont {Kardaun}},
  \bibinfo {author} {\bibfnamefont {M.}~\bibnamefont {Maslov}}, \bibinfo
  {author} {\bibfnamefont {M.}~\bibnamefont {Romanelli}}, \emph {et~al.},\
  }\bibfield  {title} {\bibinfo {title} {The updated {ITPA} global {H}-mode
  confinement database: description and analysis},\ }\href
  {https://doi.org/10.1088/1741-4326/abdb91} {\bibfield  {journal} {\bibinfo
  {journal} {Nucl. Fusion}\ }\textbf {\bibinfo {volume} {61}},\ \bibinfo
  {pages} {076006} (\bibinfo {year} {2021})}\BibitemShut {NoStop}%
\bibitem [{\citenamefont {Wade}\ and\ \citenamefont {Leuer}(2021)}]{wade2021}%
  \BibitemOpen
  \bibfield  {author} {\bibinfo {author} {\bibfnamefont {M.~R.}\ \bibnamefont
  {Wade}}\ and\ \bibinfo {author} {\bibfnamefont {J.~A.}\ \bibnamefont
  {Leuer}},\ }\bibfield  {title} {\bibinfo {title} {Cost drivers for a
  tokamak-based compact pilot plant},\ }\href
  {https://doi.org/10.1080/15361055.2020.1858670} {\bibfield  {journal}
  {\bibinfo  {journal} {Fusion Sci. Technol.}\ }\textbf {\bibinfo {volume}
  {77}},\ \bibinfo {pages} {119} (\bibinfo {year} {2021})}\BibitemShut
  {NoStop}%
\bibitem [{\citenamefont {Schuster}\ \emph {et~al.}(2003)\citenamefont
  {Schuster}, \citenamefont {Krsti{\'c}},\ and\ \citenamefont
  {Tynan}}]{schuster2003}%
  \BibitemOpen
  \bibfield  {author} {\bibinfo {author} {\bibfnamefont {E.}~\bibnamefont
  {Schuster}}, \bibinfo {author} {\bibfnamefont {M.}~\bibnamefont
  {Krsti{\'c}}},\ and\ \bibinfo {author} {\bibfnamefont {G.}~\bibnamefont
  {Tynan}},\ }\bibfield  {title} {\bibinfo {title} {Burn control in fusion
  reactors via nonlinear stabilization techniques},\ }\href
  {https://doi.org/10.13182/FST03-A246} {\bibfield  {journal} {\bibinfo
  {journal} {Fusion Sci. Technol.}\ }\textbf {\bibinfo {volume} {43}},\
  \bibinfo {pages} {18} (\bibinfo {year} {2003})}\BibitemShut {NoStop}%
\bibitem [{\citenamefont {Boyer}\ and\ \citenamefont
  {Schuster}(2015)}]{boyer2015}%
  \BibitemOpen
  \bibfield  {author} {\bibinfo {author} {\bibfnamefont {M.~D.}\ \bibnamefont
  {Boyer}}\ and\ \bibinfo {author} {\bibfnamefont {E.}~\bibnamefont
  {Schuster}},\ }\bibfield  {title} {\bibinfo {title} {Nonlinear burn condition
  control in tokamaks using isotopic fuel tailoring},\ }\href
  {https://doi.org/10.1088/0029-5515/55/8/083021} {\bibfield  {journal}
  {\bibinfo  {journal} {Nucl. Fusion}\ }\textbf {\bibinfo {volume} {55}},\
  \bibinfo {pages} {083021} (\bibinfo {year} {2015})}\BibitemShut {NoStop}%
\bibitem [{\citenamefont {Humphreys}\ \emph {et~al.}(2015)\citenamefont
  {Humphreys}, \citenamefont {Ambrosino}, \citenamefont {de~Vries},
  \citenamefont {Felici}, \citenamefont {Kim}, \citenamefont {Jackson} \emph
  {et~al.}}]{humphreys2015}%
  \BibitemOpen
  \bibfield  {author} {\bibinfo {author} {\bibfnamefont {D.}~\bibnamefont
  {Humphreys}}, \bibinfo {author} {\bibfnamefont {G.}~\bibnamefont
  {Ambrosino}}, \bibinfo {author} {\bibfnamefont {P.}~\bibnamefont {de~Vries}},
  \bibinfo {author} {\bibfnamefont {F.}~\bibnamefont {Felici}}, \bibinfo
  {author} {\bibfnamefont {S.~H.}\ \bibnamefont {Kim}}, \bibinfo {author}
  {\bibfnamefont {G.}~\bibnamefont {Jackson}}, \emph {et~al.},\ }\bibfield
  {title} {\bibinfo {title} {Novel aspects of plasma control in {ITER}},\
  }\href {https://doi.org/10.1063/1.4907901} {\bibfield  {journal} {\bibinfo
  {journal} {Phys. Plasmas}\ }\textbf {\bibinfo {volume} {22}},\ \bibinfo
  {pages} {021806} (\bibinfo {year} {2015})}\BibitemShut {NoStop}%
\bibitem [{\citenamefont {Strait}\ \emph {et~al.}(2019)\citenamefont {Strait},
  \citenamefont {Barr}, \citenamefont {Baruzzo}, \citenamefont {Berkery},
  \citenamefont {Buttery} \emph {et~al.}}]{strait}%
  \BibitemOpen
  \bibfield  {author} {\bibinfo {author} {\bibfnamefont {E.~J.}\ \bibnamefont
  {Strait}}, \bibinfo {author} {\bibfnamefont {J.~L.}\ \bibnamefont {Barr}},
  \bibinfo {author} {\bibfnamefont {M.}~\bibnamefont {Baruzzo}}, \bibinfo
  {author} {\bibfnamefont {J.~W.}\ \bibnamefont {Berkery}}, \bibinfo {author}
  {\bibfnamefont {R.~J.}\ \bibnamefont {Buttery}}, \emph {et~al.},\ }\bibfield
  {title} {\bibinfo {title} {Progress in disruption prevention for {ITER}},\
  }\href {https://doi.org/10.1088/1741-4326/ab15de} {\bibfield  {journal}
  {\bibinfo  {journal} {Nucl. Fusion}\ }\textbf {\bibinfo {volume} {59}},\
  \bibinfo {pages} {112012} (\bibinfo {year} {2019})}\BibitemShut {NoStop}%
\bibitem [{\citenamefont {de~Vries}\ \emph {et~al.}(2018)\citenamefont
  {de~Vries}, \citenamefont {Luce}, \citenamefont {Bae}, \citenamefont
  {Gerhardt}, \citenamefont {Gong}, \citenamefont {Gribov} \emph
  {et~al.}}]{devries2018}%
  \BibitemOpen
  \bibfield  {author} {\bibinfo {author} {\bibfnamefont {P.~C.}\ \bibnamefont
  {de~Vries}}, \bibinfo {author} {\bibfnamefont {T.~C.}\ \bibnamefont {Luce}},
  \bibinfo {author} {\bibfnamefont {Y.~S.}\ \bibnamefont {Bae}}, \bibinfo
  {author} {\bibfnamefont {S.}~\bibnamefont {Gerhardt}}, \bibinfo {author}
  {\bibfnamefont {X.}~\bibnamefont {Gong}}, \bibinfo {author} {\bibfnamefont
  {Y.}~\bibnamefont {Gribov}}, \emph {et~al.},\ }\bibfield  {title} {\bibinfo
  {title} {Multi-machine analysis of termination scenarios with comparison to
  simulations of controlled shutdown of {ITER} discharges},\ }\href
  {https://doi.org/10.1088/1741-4326/aa9c4c} {\bibfield  {journal} {\bibinfo
  {journal} {Nucl. Fusion}\ }\textbf {\bibinfo {volume} {58}},\ \bibinfo
  {pages} {026019} (\bibinfo {year} {2018})}\BibitemShut {NoStop}%
\bibitem [{\citenamefont {Yang}\ \emph {et~al.}(2013)\citenamefont {Yang},
  \citenamefont {Maruyama}, \citenamefont {Kiss}, \citenamefont {Ciattaglia},
  \citenamefont {Putvinski}, \citenamefont {Yoshino} \emph {et~al.}}]{fpss}%
  \BibitemOpen
  \bibfield  {author} {\bibinfo {author} {\bibfnamefont {Y.}~\bibnamefont
  {Yang}}, \bibinfo {author} {\bibfnamefont {S.}~\bibnamefont {Maruyama}},
  \bibinfo {author} {\bibfnamefont {G.}~\bibnamefont {Kiss}}, \bibinfo {author}
  {\bibfnamefont {S.}~\bibnamefont {Ciattaglia}}, \bibinfo {author}
  {\bibfnamefont {S.}~\bibnamefont {Putvinski}}, \bibinfo {author}
  {\bibfnamefont {R.}~\bibnamefont {Yoshino}}, \emph {et~al.},\ }\bibfield
  {title} {\bibinfo {title} {Concept design of fusion power shutdown system for
  {ITER}},\ }\href {https://doi.org/10.1016/j.fusengdes.2013.01.055} {\bibfield
   {journal} {\bibinfo  {journal} {Fusion Eng. Des.}\ }\textbf {\bibinfo
  {volume} {88}},\ \bibinfo {pages} {824} (\bibinfo {year} {2013})}\BibitemShut
  {NoStop}%
\bibitem [{\citenamefont {Kulsrud}\ \emph {et~al.}(1982)\citenamefont
  {Kulsrud}, \citenamefont {Furth}, \citenamefont {Valeo},\ and\ \citenamefont
  {Goldhaber}}]{kulsrud}%
  \BibitemOpen
  \bibfield  {author} {\bibinfo {author} {\bibfnamefont {R.~M.}\ \bibnamefont
  {Kulsrud}}, \bibinfo {author} {\bibfnamefont {H.~P.}\ \bibnamefont {Furth}},
  \bibinfo {author} {\bibfnamefont {E.~J.}\ \bibnamefont {Valeo}},\ and\
  \bibinfo {author} {\bibfnamefont {M.}~\bibnamefont {Goldhaber}},\ }\bibfield
  {title} {\bibinfo {title} {Fusion reactor plasmas with polarized nuclei},\
  }\href {https://doi.org/10.1103/PhysRevLett.49.1248} {\bibfield  {journal}
  {\bibinfo  {journal} {Phys. Rev. Lett.}\ }\textbf {\bibinfo {volume} {49}},\
  \bibinfo {pages} {1248} (\bibinfo {year} {1982})}\BibitemShut {NoStop}%
\bibitem [{\citenamefont {Cowley}\ \emph {et~al.}(1986)\citenamefont {Cowley},
  \citenamefont {Kulsrud},\ and\ \citenamefont {Valeo}}]{cowley1986}%
  \BibitemOpen
  \bibfield  {author} {\bibinfo {author} {\bibfnamefont {S.~C.}\ \bibnamefont
  {Cowley}}, \bibinfo {author} {\bibfnamefont {R.~M.}\ \bibnamefont
  {Kulsrud}},\ and\ \bibinfo {author} {\bibfnamefont {E.}~\bibnamefont
  {Valeo}},\ }\bibfield  {title} {\bibinfo {title} {A kinetic equation for
  spin-polarized plasmas},\ }\href {https://doi.org/10.1063/1.865726}
  {\bibfield  {journal} {\bibinfo  {journal} {Phys. Fluids}\ }\textbf {\bibinfo
  {volume} {29}},\ \bibinfo {pages} {430} (\bibinfo {year} {1986})}\BibitemShut
  {NoStop}%
\bibitem [{\citenamefont {Hupin}\ \emph {et~al.}(2019)\citenamefont {Hupin},
  \citenamefont {Quaglioni},\ and\ \citenamefont {Navr\'atil}}]{hupin2019}%
  \BibitemOpen
  \bibfield  {author} {\bibinfo {author} {\bibfnamefont {G.}~\bibnamefont
  {Hupin}}, \bibinfo {author} {\bibfnamefont {S.}~\bibnamefont {Quaglioni}},\
  and\ \bibinfo {author} {\bibfnamefont {P.}~\bibnamefont {Navr\'atil}},\
  }\bibfield  {title} {\bibinfo {title} {Ab initio predictions for polarized
  deuterium-tritium thermonuclear fusion},\ }\href
  {https://doi.org/10.1038/s41467-018-08052-6} {\bibfield  {journal} {\bibinfo
  {journal} {Nat. Commun.}\ }\textbf {\bibinfo {volume} {10}},\ \bibinfo
  {pages} {351} (\bibinfo {year} {2019})}\BibitemShut {NoStop}%
\bibitem [{\citenamefont {More}(1983)}]{more1983}%
  \BibitemOpen
  \bibfield  {author} {\bibinfo {author} {\bibfnamefont {R.~M.}\ \bibnamefont
  {More}},\ }\bibfield  {title} {\bibinfo {title} {Nuclear spin-polarized fuel
  in inertial fusion},\ }\href {https://doi.org/10.1103/PhysRevLett.51.396}
  {\bibfield  {journal} {\bibinfo  {journal} {Phys. Rev. Lett.}\ }\textbf
  {\bibinfo {volume} {51}},\ \bibinfo {pages} {396} (\bibinfo {year}
  {1983})}\BibitemShut {NoStop}%
\bibitem [{\citenamefont {Coppi}\ \emph {et~al.}(1983)\citenamefont {Coppi},
  \citenamefont {Pegoraro},\ and\ \citenamefont {Ramos}}]{coppi1983}%
  \BibitemOpen
  \bibfield  {author} {\bibinfo {author} {\bibfnamefont {B.}~\bibnamefont
  {Coppi}}, \bibinfo {author} {\bibfnamefont {F.}~\bibnamefont {Pegoraro}},\
  and\ \bibinfo {author} {\bibfnamefont {J.~J.}\ \bibnamefont {Ramos}},\
  }\bibfield  {title} {\bibinfo {title} {Instability of fusing plasmas and
  spin-depolarization processes},\ }\href
  {https://doi.org/10.1103/PhysRevLett.51.892} {\bibfield  {journal} {\bibinfo
  {journal} {Phys. Rev. Lett.}\ }\textbf {\bibinfo {volume} {51}},\ \bibinfo
  {pages} {892} (\bibinfo {year} {1983})}\BibitemShut {NoStop}%
\bibitem [{\citenamefont {Micklich}\ and\ \citenamefont
  {Jassby}(1984)}]{micklich1984}%
  \BibitemOpen
  \bibfield  {author} {\bibinfo {author} {\bibfnamefont {B.~J.}\ \bibnamefont
  {Micklich}}\ and\ \bibinfo {author} {\bibfnamefont {D.~L.}\ \bibnamefont
  {Jassby}},\ }\bibfield  {title} {\bibinfo {title} {Implications of polarized
  deuterium-tritium plasmas for toroidal fusion reactors},\ }\href
  {https://doi.org/10.13182/FST84-A23090} {\bibfield  {journal} {\bibinfo
  {journal} {Nucl. Technol./Fusion}\ }\textbf {\bibinfo {volume} {5}},\
  \bibinfo {pages} {162} (\bibinfo {year} {1984})}\BibitemShut {NoStop}%
\bibitem [{\citenamefont {Kulsrud}\ \emph {et~al.}(1986)\citenamefont
  {Kulsrud}, \citenamefont {Valeo},\ and\ \citenamefont {Cowley}}]{kvc}%
  \BibitemOpen
  \bibfield  {author} {\bibinfo {author} {\bibfnamefont {R.~M.}\ \bibnamefont
  {Kulsrud}}, \bibinfo {author} {\bibfnamefont {E.~J.}\ \bibnamefont {Valeo}},\
  and\ \bibinfo {author} {\bibfnamefont {S.~C.}\ \bibnamefont {Cowley}},\
  }\bibfield  {title} {\bibinfo {title} {Physics of spin-polarized plasmas},\
  }\href {https://doi.org/10.1088/0029-5515/26/11/001} {\bibfield  {journal}
  {\bibinfo  {journal} {Nucl. Fusion}\ }\textbf {\bibinfo {volume} {26}},\
  \bibinfo {pages} {1443} (\bibinfo {year} {1986})}\BibitemShut {NoStop}%
\bibitem [{\citenamefont {{Paetz gen. Schieck}}(2010)}]{paetz2010}%
  \BibitemOpen
  \bibfield  {author} {\bibinfo {author} {\bibfnamefont {H.}~\bibnamefont
  {{Paetz gen. Schieck}}},\ }\bibfield  {title} {\bibinfo {title} {The status
  of polarized fusion},\ }\href {https://doi.org/10.1140/epja/i2010-10964-4}
  {\bibfield  {journal} {\bibinfo  {journal} {Eur. Phys. J. A}\ }\textbf
  {\bibinfo {volume} {44}},\ \bibinfo {pages} {321} (\bibinfo {year}
  {2010})}\BibitemShut {NoStop}%
\bibitem [{\citenamefont {Hu}\ \emph {et~al.}(2023)\citenamefont {Hu} \emph
  {et~al.}}]{hu2023}%
  \BibitemOpen
  \bibfield  {author} {\bibinfo {author} {\bibfnamefont {R.}~\bibnamefont {Hu}}
  \emph {et~al.},\ }\bibfield  {title} {\bibinfo {title} {Numerical study of
  spin-polarized deuterium-tritium fuel persistence in inertial confinement
  fusion implosions},\ }\href
  {https://doi.org/10.1103/PhysRevResearch.5.033115} {\bibfield  {journal}
  {\bibinfo  {journal} {Phys. Rev. Research}\ }\textbf {\bibinfo {volume}
  {5}},\ \bibinfo {pages} {033115} (\bibinfo {year} {2023})}\BibitemShut
  {NoStop}%
\bibitem [{\citenamefont {Cook}\ \emph {et~al.}(2026)\citenamefont {Cook} \emph
  {et~al.}}]{cook}%
  \BibitemOpen
  \bibfield  {author} {\bibinfo {author} {\bibfnamefont {J.~W.~S.}\
  \bibnamefont {Cook}} \emph {et~al.},\ }\bibfield  {title} {\bibinfo {title}
  {Persistence of deuterium and tritium nuclear spin-polarization in presence
  of high-frequency plasma waves},\ }\href
  {https://doi.org/10.1088/1741-4326/ae5ab8} {\bibfield  {journal} {\bibinfo
  {journal} {Nucl. Fusion}\ }\textbf {\bibinfo {volume} {66}},\ \bibinfo
  {pages} {056044} (\bibinfo {year} {2026})}\BibitemShut {NoStop}%
\bibitem [{\citenamefont {Zheng}\ \emph {et~al.}(2026)\citenamefont {Zheng}
  \emph {et~al.}}]{zheng2026}%
  \BibitemOpen
  \bibfield  {author} {\bibinfo {author} {\bibfnamefont {C.}~\bibnamefont
  {Zheng}} \emph {et~al.},\ }\bibfield  {title} {\bibinfo {title} {Preservation
  of $^3${He} ion polarization after laser-driven acceleration in plasma},\
  }\href {https://doi.org/10.1017/hpl.2026.10140} {\bibfield  {journal}
  {\bibinfo  {journal} {High Power Laser Sci. Eng.}\ }\textbf {\bibinfo
  {volume} {14}},\ \bibinfo {pages} {e48} (\bibinfo {year} {2026})}\BibitemShut
  {NoStop}%
\bibitem [{\citenamefont {Ciullo}\ \emph {et~al.}(2016)\citenamefont {Ciullo},
  \citenamefont {Engels}, \citenamefont {B\"uscher},\ and\ \citenamefont
  {Vasilyev}}]{ciullo2016}%
  \BibitemOpen
  \bibinfo {editor} {\bibfnamefont {G.}~\bibnamefont {Ciullo}}, \bibinfo
  {editor} {\bibfnamefont {R.}~\bibnamefont {Engels}}, \bibinfo {editor}
  {\bibfnamefont {M.}~\bibnamefont {B\"uscher}},\ and\ \bibinfo {editor}
  {\bibfnamefont {A.}~\bibnamefont {Vasilyev}},\ eds.,\ \href
  {https://doi.org/10.1007/978-3-319-39471-8} {\emph {\bibinfo {title} {Nuclear
  Fusion with Polarized Fuel}}},\ \bibinfo {series} {Springer Proceedings in
  Physics}, Vol.\ \bibinfo {volume} {187}\ (\bibinfo  {publisher} {Springer},\
  \bibinfo {address} {Cham},\ \bibinfo {year} {2016})\BibitemShut {NoStop}%
\bibitem [{\citenamefont {Spiliotis}\ \emph {et~al.}(2021)\citenamefont
  {Spiliotis} \emph {et~al.}}]{spiliotis2021}%
  \BibitemOpen
  \bibfield  {author} {\bibinfo {author} {\bibfnamefont {A.~K.}\ \bibnamefont
  {Spiliotis}} \emph {et~al.},\ }\bibfield  {title} {\bibinfo {title}
  {Ultrahigh-density spin-polarized hydrogen isotopes from the
  photodissociation of hydrogen halides: new applications for laser-ion
  acceleration, magnetometry, and polarized nuclear fusion},\ }\href
  {https://doi.org/10.1038/s41377-021-00476-y} {\bibfield  {journal} {\bibinfo
  {journal} {Light Sci. Appl.}\ }\textbf {\bibinfo {volume} {10}},\ \bibinfo
  {pages} {35} (\bibinfo {year} {2021})}\BibitemShut {NoStop}%
\bibitem [{\citenamefont {Kannis}\ and\ \citenamefont
  {Rakitzis}(2021)}]{kannis2021}%
  \BibitemOpen
  \bibfield  {author} {\bibinfo {author} {\bibfnamefont {C.~S.}\ \bibnamefont
  {Kannis}}\ and\ \bibinfo {author} {\bibfnamefont {T.~P.}\ \bibnamefont
  {Rakitzis}},\ }\bibfield  {title} {\bibinfo {title} {Macroscopic production
  of highly nuclear-spin-polarized molecules from {IR}-excitation and
  photodissociation of molecular beams},\ }\href
  {https://doi.org/10.1016/j.cplett.2021.139092} {\bibfield  {journal}
  {\bibinfo  {journal} {Chem. Phys. Lett.}\ }\textbf {\bibinfo {volume}
  {784}},\ \bibinfo {pages} {139092} (\bibinfo {year} {2021})}\BibitemShut
  {NoStop}%
\bibitem [{\citenamefont {Ciullo}\ \emph {et~al.}(2022)\citenamefont {Ciullo}
  \emph {et~al.}}]{ciullo2022}%
  \BibitemOpen
  \bibfield  {author} {\bibinfo {author} {\bibfnamefont {G.}~\bibnamefont
  {Ciullo}} \emph {et~al.},\ }\bibfield  {title} {\bibinfo {title}
  {Polarization {RE}search for {F}usion {E}xperiments and {R}eactors: the
  {PREFER} collaboration, purposes and present status},\ }\href
  {https://doi.org/10.7566/JPSCP.37.021302} {\bibfield  {journal} {\bibinfo
  {journal} {JPS Conf. Proc.}\ }\textbf {\bibinfo {volume} {37}},\ \bibinfo
  {pages} {021302} (\bibinfo {year} {2022})}\BibitemShut {NoStop}%
\bibitem [{\citenamefont {Kannis}\ and\ \citenamefont
  {Rakitzis}(2026)}]{kannis2026}%
  \BibitemOpen
  \bibfield  {author} {\bibinfo {author} {\bibfnamefont {C.~S.}\ \bibnamefont
  {Kannis}}\ and\ \bibinfo {author} {\bibfnamefont {T.~P.}\ \bibnamefont
  {Rakitzis}},\ }\bibfield  {title} {\bibinfo {title} {Production of
  spin-polarized molecular beams via microwave or infrared rotational
  excitation},\ }\href {https://doi.org/10.1016/j.chemphys.2026.113200}
  {\bibfield  {journal} {\bibinfo  {journal} {Chem. Phys.}\ }\textbf {\bibinfo
  {volume} {607}},\ \bibinfo {pages} {113200} (\bibinfo {year}
  {2026})}\BibitemShut {NoStop}%
\bibitem [{\citenamefont {Baylor}\ \emph {et~al.}(2023)\citenamefont {Baylor}
  \emph {et~al.}}]{baylor}%
  \BibitemOpen
  \bibfield  {author} {\bibinfo {author} {\bibfnamefont {L.~R.}\ \bibnamefont
  {Baylor}} \emph {et~al.},\ }\bibfield  {title} {\bibinfo {title} {Polarized
  fusion and potential in situ tests of fuel polarization survival in a tokamak
  plasma},\ }\href {https://doi.org/10.1088/1741-4326/acc3ae} {\bibfield
  {journal} {\bibinfo  {journal} {Nucl. Fusion}\ }\textbf {\bibinfo {volume}
  {63}},\ \bibinfo {pages} {076009} (\bibinfo {year} {2023})}\BibitemShut
  {NoStop}%
\bibitem [{\citenamefont {Garcia}\ \emph {et~al.}(2023)\citenamefont {Garcia},
  \citenamefont {Heidbrink},\ and\ \citenamefont {Sandorfi}}]{garcia2023}%
  \BibitemOpen
  \bibfield  {author} {\bibinfo {author} {\bibfnamefont {A.~V.}\ \bibnamefont
  {Garcia}}, \bibinfo {author} {\bibfnamefont {W.~W.}\ \bibnamefont
  {Heidbrink}},\ and\ \bibinfo {author} {\bibfnamefont {A.~M.}\ \bibnamefont
  {Sandorfi}},\ }\bibfield  {title} {\bibinfo {title} {Conceptual design of
  {DIII-D} experiments to diagnose the lifetime of spin polarized fuel},\
  }\href {https://doi.org/10.1088/1741-4326/acaf0d} {\bibfield  {journal}
  {\bibinfo  {journal} {Nucl. Fusion}\ }\textbf {\bibinfo {volume} {63}},\
  \bibinfo {pages} {026030} (\bibinfo {year} {2023})}\BibitemShut {NoStop}%
\bibitem [{\citenamefont {Heidbrink}\ \emph {et~al.}(2024)\citenamefont
  {Heidbrink} \emph {et~al.}}]{heidbrink2024}%
  \BibitemOpen
  \bibfield  {author} {\bibinfo {author} {\bibfnamefont {W.~W.}\ \bibnamefont
  {Heidbrink}} \emph {et~al.},\ }\bibfield  {title} {\bibinfo {title} {A
  research program to measure the lifetime of spin polarized fuel},\ }\href
  {https://doi.org/10.3389/fphy.2024.1355212} {\bibfield  {journal} {\bibinfo
  {journal} {Front. Phys.}\ }\textbf {\bibinfo {volume} {12}},\ \bibinfo
  {pages} {1355212} (\bibinfo {year} {2024})}\BibitemShut {NoStop}%
\bibitem [{\citenamefont {Garcia}\ \emph {et~al.}(2025)\citenamefont {Garcia},
  \citenamefont {Heidbrink},\ and\ \citenamefont {Sandorfi}}]{garcia2025}%
  \BibitemOpen
  \bibfield  {author} {\bibinfo {author} {\bibfnamefont {A.}~\bibnamefont
  {Garcia}}, \bibinfo {author} {\bibfnamefont {W.~W.}\ \bibnamefont
  {Heidbrink}},\ and\ \bibinfo {author} {\bibfnamefont {A.~M.}\ \bibnamefont
  {Sandorfi}},\ }\bibfield  {title} {\bibinfo {title} {The use of {D-D}
  reactions to diagnose the lifetime of spin polarized fuel},\ }\href
  {https://doi.org/10.1088/1741-4326/adb59b} {\bibfield  {journal} {\bibinfo
  {journal} {Nucl. Fusion}\ }\textbf {\bibinfo {volume} {65}},\ \bibinfo
  {pages} {046005} (\bibinfo {year} {2025})}\BibitemShut {NoStop}%
\bibitem [{\citenamefont {Parisi}\ \emph {et~al.}(2024)\citenamefont {Parisi},
  \citenamefont {Diallo},\ and\ \citenamefont {Schwartz}}]{parisi2024}%
  \BibitemOpen
  \bibfield  {author} {\bibinfo {author} {\bibfnamefont {J.~F.}\ \bibnamefont
  {Parisi}}, \bibinfo {author} {\bibfnamefont {A.}~\bibnamefont {Diallo}},\
  and\ \bibinfo {author} {\bibfnamefont {J.~A.}\ \bibnamefont {Schwartz}},\
  }\bibfield  {title} {\bibinfo {title} {Simultaneous enhancement of tritium
  burn efficiency and fusion power with low-tritium spin-polarized fuel},\
  }\href {https://doi.org/10.1088/1741-4326/ad7da3} {\bibfield  {journal}
  {\bibinfo  {journal} {Nucl. Fusion}\ }\textbf {\bibinfo {volume} {64}},\
  \bibinfo {pages} {126019} (\bibinfo {year} {2024})}\BibitemShut {NoStop}%
\bibitem [{\citenamefont {Bae}\ \emph {et~al.}(2025)\citenamefont {Bae} \emph
  {et~al.}}]{bae2025}%
  \BibitemOpen
  \bibfield  {author} {\bibinfo {author} {\bibfnamefont {J.~W.}\ \bibnamefont
  {Bae}} \emph {et~al.},\ }\bibfield  {title} {\bibinfo {title} {Neutronics
  analysis of spin-polarized fuel in spherical tokamaks},\ }\href
  {https://doi.org/10.1088/1741-4326/adf3c6} {\bibfield  {journal} {\bibinfo
  {journal} {Nucl. Fusion}\ }\textbf {\bibinfo {volume} {65}},\ \bibinfo
  {pages} {086051} (\bibinfo {year} {2025})}\BibitemShut {NoStop}%
\bibitem [{\citenamefont {Parisi}\ and\ \citenamefont
  {Diallo}(2025)}]{parisi2025power}%
  \BibitemOpen
  \bibfield  {author} {\bibinfo {author} {\bibfnamefont {J.~F.}\ \bibnamefont
  {Parisi}}\ and\ \bibinfo {author} {\bibfnamefont {A.}~\bibnamefont
  {Diallo}},\ }\href@noop {} {\bibinfo {title} {Electric power enhancement
  using spin-polarized fuel in fusion power plants}} (\bibinfo {year} {2025}),\
  \Eprint {https://arxiv.org/abs/2502.15941} {arXiv:2502.15941} \BibitemShut
  {NoStop}%
\bibitem [{\citenamefont {Parisi}\ \emph {et~al.}(2025)\citenamefont {Parisi},
  \citenamefont {Diallo},\ and\ \citenamefont {Meschini}}]{parisi2025dhe3}%
  \BibitemOpen
  \bibfield  {author} {\bibinfo {author} {\bibfnamefont {J.~F.}\ \bibnamefont
  {Parisi}}, \bibinfo {author} {\bibfnamefont {A.}~\bibnamefont {Diallo}},\
  and\ \bibinfo {author} {\bibfnamefont {S.}~\bibnamefont {Meschini}},\
  }\href@noop {} {\bibinfo {title} {Revisiting fusion in {D-$^3$He} plasmas
  with spin-polarized fuel}} (\bibinfo {year} {2025}),\ \Eprint
  {https://arxiv.org/abs/2504.09869} {arXiv:2504.09869} \BibitemShut {NoStop}%
\bibitem [{\citenamefont {Parisi}\ \emph {et~al.}(2026)\citenamefont {Parisi},
  \citenamefont {Diallo},\ and\ \citenamefont {Cook}}]{parisi2026channel}%
  \BibitemOpen
  \bibfield  {author} {\bibinfo {author} {\bibfnamefont {J.~F.}\ \bibnamefont
  {Parisi}}, \bibinfo {author} {\bibfnamefont {A.}~\bibnamefont {Diallo}},\
  and\ \bibinfo {author} {\bibfnamefont {J.~W.~S.}\ \bibnamefont {Cook}},\
  }\href@noop {} {\bibinfo {title} {Enhanced alpha channeling with
  spin-polarized fuel}} (\bibinfo {year} {2026}),\ \Eprint
  {https://arxiv.org/abs/2607.25313} {arXiv:2607.25313} \BibitemShut {NoStop}%
\bibitem [{\citenamefont {Mitarai}\ \emph {et~al.}(1992)\citenamefont
  {Mitarai}, \citenamefont {Hasuyama},\ and\ \citenamefont {Wakuta}}]{mitarai}%
  \BibitemOpen
  \bibfield  {author} {\bibinfo {author} {\bibfnamefont {O.}~\bibnamefont
  {Mitarai}}, \bibinfo {author} {\bibfnamefont {H.}~\bibnamefont {Hasuyama}},\
  and\ \bibinfo {author} {\bibfnamefont {Y.}~\bibnamefont {Wakuta}},\
  }\bibfield  {title} {\bibinfo {title} {Spin polarization effect on ignition
  access condition for {D-T} and {D-$^3$He} tokamak fusion reactors},\ }\href
  {https://doi.org/10.13182/FST92-A29720} {\bibfield  {journal} {\bibinfo
  {journal} {Fusion Technol.}\ }\textbf {\bibinfo {volume} {21}},\ \bibinfo
  {pages} {2265} (\bibinfo {year} {1992})}\BibitemShut {NoStop}%
\bibitem [{\citenamefont {Temporal}\ \emph {et~al.}(2012)\citenamefont
  {Temporal} \emph {et~al.}}]{temporal2012}%
  \BibitemOpen
  \bibfield  {author} {\bibinfo {author} {\bibfnamefont {M.}~\bibnamefont
  {Temporal}} \emph {et~al.},\ }\bibfield  {title} {\bibinfo {title} {Ignition
  conditions for inertial confinement fusion targets with a nuclear
  spin-polarized {DT} fuel},\ }\href
  {https://doi.org/10.1088/0029-5515/52/10/103011} {\bibfield  {journal}
  {\bibinfo  {journal} {Nucl. Fusion}\ }\textbf {\bibinfo {volume} {52}},\
  \bibinfo {pages} {103011} (\bibinfo {year} {2012})}\BibitemShut {NoStop}%
\bibitem [{\citenamefont {Delgado-Aparicio}\ \emph {et~al.}(2026)\citenamefont
  {Delgado-Aparicio}, \citenamefont {Ono},\ and\ \citenamefont
  {Menard}}]{lfdav}%
  \BibitemOpen
  \bibfield  {author} {\bibinfo {author} {\bibfnamefont {L.~F.}\ \bibnamefont
  {Delgado-Aparicio}}, \bibinfo {author} {\bibfnamefont {M.}~\bibnamefont
  {Ono}},\ and\ \bibinfo {author} {\bibfnamefont {J.~E.}\ \bibnamefont
  {Menard}},\ }\bibfield  {title} {\bibinfo {title} {Generalized
  {Lawson-Cordey-Mills} accessibility of fusion ignition},\ }\href
  {https://doi.org/10.1103/mmc9-nzfx} {\bibfield  {journal} {\bibinfo
  {journal} {Phys. Rev. Lett.}\ }\textbf {\bibinfo {volume} {137}},\ \bibinfo
  {pages} {115102} (\bibinfo {year} {2026})}\BibitemShut {NoStop}%
\bibitem [{\citenamefont {Abdou}\ \emph {et~al.}(2021)\citenamefont {Abdou}
  \emph {et~al.}}]{abdou}%
  \BibitemOpen
  \bibfield  {author} {\bibinfo {author} {\bibfnamefont {M.}~\bibnamefont
  {Abdou}} \emph {et~al.},\ }\bibfield  {title} {\bibinfo {title} {Physics and
  technology considerations for the deuterium-tritium fuel cycle and conditions
  for tritium fuel self sufficiency},\ }\href
  {https://doi.org/10.1088/1741-4326/abbf35} {\bibfield  {journal} {\bibinfo
  {journal} {Nucl. Fusion}\ }\textbf {\bibinfo {volume} {61}},\ \bibinfo
  {pages} {013001} (\bibinfo {year} {2021})}\BibitemShut {NoStop}%
\bibitem [{\citenamefont {Whyte}\ \emph {et~al.}(2023)\citenamefont {Whyte},
  \citenamefont {Delaporte-Mathurin}, \citenamefont {Ferry},\ and\
  \citenamefont {Meschini}}]{whyte2023}%
  \BibitemOpen
  \bibfield  {author} {\bibinfo {author} {\bibfnamefont {D.~G.}\ \bibnamefont
  {Whyte}}, \bibinfo {author} {\bibfnamefont {R.}~\bibnamefont
  {Delaporte-Mathurin}}, \bibinfo {author} {\bibfnamefont {S.~E.}\ \bibnamefont
  {Ferry}},\ and\ \bibinfo {author} {\bibfnamefont {S.}~\bibnamefont
  {Meschini}},\ }\bibfield  {title} {\bibinfo {title} {Tritium burn efficiency
  in deuterium-tritium magnetic fusion},\ }\href
  {https://doi.org/10.1088/1741-4326/acf3fb} {\bibfield  {journal} {\bibinfo
  {journal} {Nucl. Fusion}\ }\textbf {\bibinfo {volume} {63}},\ \bibinfo
  {pages} {126019} (\bibinfo {year} {2023})}\BibitemShut {NoStop}%
\bibitem [{\citenamefont {Takahashi}\ and\ \citenamefont
  {Nakao}(1999)}]{takahashi}%
  \BibitemOpen
  \bibfield  {author} {\bibinfo {author} {\bibfnamefont {T.}~\bibnamefont
  {Takahashi}}\ and\ \bibinfo {author} {\bibfnamefont {Y.}~\bibnamefont
  {Nakao}},\ }\href {https://www.nifs.ac.jp/report/NIFS-592.pdf} {\emph
  {\bibinfo {title} {Thermonuclear Reactivity of {D-T} Fusion Plasma with
  Spin-Polarized Fuel}}},\ \bibinfo {type} {Tech. Rep.}\ \bibinfo {number}
  {NIFS-592}\ (\bibinfo  {institution} {National Institute for Fusion
  Science},\ \bibinfo {year} {1999})\BibitemShut {NoStop}%
\bibitem [{\citenamefont {Yamada}\ \emph {et~al.}(2005)\citenamefont {Yamada}
  \emph {et~al.}}]{iss04}%
  \BibitemOpen
  \bibfield  {author} {\bibinfo {author} {\bibfnamefont {H.}~\bibnamefont
  {Yamada}} \emph {et~al.},\ }\bibfield  {title} {\bibinfo {title}
  {Characterization of energy confinement in net-current free plasmas using the
  extended international stellarator database},\ }\href
  {https://doi.org/10.1088/0029-5515/45/12/024} {\bibfield  {journal} {\bibinfo
   {journal} {Nucl. Fusion}\ }\textbf {\bibinfo {volume} {45}},\ \bibinfo
  {pages} {1684} (\bibinfo {year} {2005})}\BibitemShut {NoStop}%
\bibitem [{\citenamefont {P{\"u}tterich}\ \emph {et~al.}(2010)\citenamefont
  {P{\"u}tterich} \emph {et~al.}}]{putterich}%
  \BibitemOpen
  \bibfield  {author} {\bibinfo {author} {\bibfnamefont {T.}~\bibnamefont
  {P{\"u}tterich}} \emph {et~al.},\ }\bibfield  {title} {\bibinfo {title}
  {Calculation and experimental test of the cooling factor of tungsten},\
  }\href {https://doi.org/10.1088/0029-5515/50/2/025012} {\bibfield  {journal}
  {\bibinfo  {journal} {Nucl. Fusion}\ }\textbf {\bibinfo {volume} {50}},\
  \bibinfo {pages} {025012} (\bibinfo {year} {2010})}\BibitemShut {NoStop}%
\bibitem [{\citenamefont {Citrin}\ \emph {et~al.}(2024)\citenamefont {Citrin}
  \emph {et~al.}}]{torax}%
  \BibitemOpen
  \bibfield  {author} {\bibinfo {author} {\bibfnamefont {J.}~\bibnamefont
  {Citrin}} \emph {et~al.},\ }\href@noop {} {\bibinfo {title} {{TORAX}: A fast
  and differentiable tokamak transport simulator in {JAX}}} (\bibinfo {year}
  {2024}),\ \Eprint {https://arxiv.org/abs/2406.06718} {arXiv:2406.06718}
  \BibitemShut {NoStop}%
\bibitem [{\citenamefont {Sorbom}\ \emph {et~al.}(2015)\citenamefont {Sorbom},
  \citenamefont {Ball}, \citenamefont {Palmer}, \citenamefont {Mangiarotti},
  \citenamefont {Sierchio} \emph {et~al.}}]{sorbom2015}%
  \BibitemOpen
  \bibfield  {author} {\bibinfo {author} {\bibfnamefont {B.~N.}\ \bibnamefont
  {Sorbom}}, \bibinfo {author} {\bibfnamefont {J.}~\bibnamefont {Ball}},
  \bibinfo {author} {\bibfnamefont {T.~R.}\ \bibnamefont {Palmer}}, \bibinfo
  {author} {\bibfnamefont {F.~J.}\ \bibnamefont {Mangiarotti}}, \bibinfo
  {author} {\bibfnamefont {J.~M.}\ \bibnamefont {Sierchio}}, \emph {et~al.},\
  }\bibfield  {title} {\bibinfo {title} {{ARC}: A compact, high-field, fusion
  nuclear science facility and demonstration power plant with demountable
  magnets},\ }\href {https://doi.org/10.1016/j.fusengdes.2015.07.008}
  {\bibfield  {journal} {\bibinfo  {journal} {Fusion Eng. Des.}\ }\textbf
  {\bibinfo {volume} {100}},\ \bibinfo {pages} {378} (\bibinfo {year}
  {2015})}\BibitemShut {NoStop}%
\bibitem [{\citenamefont {Abragam}(1961)}]{abragam}%
  \BibitemOpen
  \bibfield  {author} {\bibinfo {author} {\bibfnamefont {A.}~\bibnamefont
  {Abragam}},\ }\href@noop {} {\emph {\bibinfo {title} {The Principles of
  Nuclear Magnetism}}}\ (\bibinfo  {publisher} {Oxford University Press},\
  \bibinfo {address} {Oxford},\ \bibinfo {year} {1961})\BibitemShut {NoStop}%
\bibitem [{\citenamefont {Lamalle}\ \emph {et~al.}(2013)\citenamefont
  {Lamalle}, \citenamefont {Beaumont}, \citenamefont {Kazarian}, \citenamefont
  {Gassmann}, \citenamefont {Agarici} \emph {et~al.}}]{lamalle2013}%
  \BibitemOpen
  \bibfield  {author} {\bibinfo {author} {\bibfnamefont {P.}~\bibnamefont
  {Lamalle}}, \bibinfo {author} {\bibfnamefont {B.}~\bibnamefont {Beaumont}},
  \bibinfo {author} {\bibfnamefont {F.}~\bibnamefont {Kazarian}}, \bibinfo
  {author} {\bibfnamefont {T.}~\bibnamefont {Gassmann}}, \bibinfo {author}
  {\bibfnamefont {G.}~\bibnamefont {Agarici}}, \emph {et~al.},\ }\bibfield
  {title} {\bibinfo {title} {Status of the {ITER} ion cyclotron {H\&CD}
  system},\ }\href {https://doi.org/10.1016/j.fusengdes.2012.11.027} {\bibfield
   {journal} {\bibinfo  {journal} {Fusion Eng. Des.}\ }\textbf {\bibinfo
  {volume} {88}},\ \bibinfo {pages} {517} (\bibinfo {year} {2013})}\BibitemShut
  {NoStop}%
\bibitem [{\citenamefont {Lin}\ \emph {et~al.}(2020)\citenamefont {Lin},
  \citenamefont {Wright},\ and\ \citenamefont {Wukitch}}]{lin2020}%
  \BibitemOpen
  \bibfield  {author} {\bibinfo {author} {\bibfnamefont {Y.}~\bibnamefont
  {Lin}}, \bibinfo {author} {\bibfnamefont {J.~C.}\ \bibnamefont {Wright}},\
  and\ \bibinfo {author} {\bibfnamefont {S.~J.}\ \bibnamefont {Wukitch}},\
  }\bibfield  {title} {\bibinfo {title} {Physics basis for the {ICRF} system of
  the {SPARC} tokamak},\ }\href {https://doi.org/10.1017/S0022377820001269}
  {\bibfield  {journal} {\bibinfo  {journal} {J. Plasma Phys.}\ }\textbf
  {\bibinfo {volume} {86}},\ \bibinfo {pages} {865860506} (\bibinfo {year}
  {2020})}\BibitemShut {NoStop}%
\bibitem [{\citenamefont {Brambilla}(1999)}]{brambilla1999}%
  \BibitemOpen
  \bibfield  {author} {\bibinfo {author} {\bibfnamefont {M.}~\bibnamefont
  {Brambilla}},\ }\bibfield  {title} {\bibinfo {title} {Numerical simulation of
  ion cyclotron waves in tokamak plasmas},\ }\href
  {https://doi.org/10.1088/0741-3335/41/1/002} {\bibfield  {journal} {\bibinfo
  {journal} {Plasma Phys. Control. Fusion}\ }\textbf {\bibinfo {volume} {41}},\
  \bibinfo {pages} {1} (\bibinfo {year} {1999})}\BibitemShut {NoStop}%
\bibitem [{\citenamefont {Jaeger}\ \emph {et~al.}(2001)\citenamefont {Jaeger},
  \citenamefont {Berry}, \citenamefont {D'Azevedo}, \citenamefont {Batchelor},\
  and\ \citenamefont {Carter}}]{jaeger2001}%
  \BibitemOpen
  \bibfield  {author} {\bibinfo {author} {\bibfnamefont {E.~F.}\ \bibnamefont
  {Jaeger}}, \bibinfo {author} {\bibfnamefont {L.~A.}\ \bibnamefont {Berry}},
  \bibinfo {author} {\bibfnamefont {E.}~\bibnamefont {D'Azevedo}}, \bibinfo
  {author} {\bibfnamefont {D.~B.}\ \bibnamefont {Batchelor}},\ and\ \bibinfo
  {author} {\bibfnamefont {M.~D.}\ \bibnamefont {Carter}},\ }\bibfield  {title}
  {\bibinfo {title} {All-orders spectral calculation of radio-frequency heating
  in two-dimensional toroidal plasmas},\ }\href
  {https://doi.org/10.1063/1.1359516} {\bibfield  {journal} {\bibinfo
  {journal} {Phys. Plasmas}\ }\textbf {\bibinfo {volume} {8}},\ \bibinfo
  {pages} {1573} (\bibinfo {year} {2001})}\BibitemShut {NoStop}%
\bibitem [{\citenamefont {Sheffield}\ and\ \citenamefont
  {Milora}(2016)}]{sheffield2016}%
  \BibitemOpen
  \bibfield  {author} {\bibinfo {author} {\bibfnamefont {J.}~\bibnamefont
  {Sheffield}}\ and\ \bibinfo {author} {\bibfnamefont {S.~L.}\ \bibnamefont
  {Milora}},\ }\bibfield  {title} {\bibinfo {title} {Generic magnetic fusion
  reactor revisited},\ }\href {https://doi.org/10.13182/FST15-157} {\bibfield
  {journal} {\bibinfo  {journal} {Fusion Sci. Technol.}\ }\textbf {\bibinfo
  {volume} {70}},\ \bibinfo {pages} {14} (\bibinfo {year} {2016})}\BibitemShut
  {NoStop}%
\bibitem [{\citenamefont {Di~Nicola}\ and\ \citenamefont
  {Pedretti}(1987)}]{dinicola1987}%
  \BibitemOpen
  \bibfield  {author} {\bibinfo {author} {\bibfnamefont {C.}~\bibnamefont
  {Di~Nicola}}\ and\ \bibinfo {author} {\bibfnamefont {E.}~\bibnamefont
  {Pedretti}},\ }\bibfield  {title} {\bibinfo {title} {Uncollided neutron
  fluences in a mirror-like machine ({MFTF-B}) with polarized {DT} plasmas},\
  }in\ \href {https://doi.org/10.1007/978-1-4757-5930-3_16} {\emph {\bibinfo
  {booktitle} {Muon-Catalyzed Fusion and Fusion with Polarized Nuclei}}},\
  \bibinfo {editor} {edited by\ \bibinfo {editor} {\bibfnamefont
  {B.}~\bibnamefont {Brunelli}}\ and\ \bibinfo {editor} {\bibfnamefont {G.~G.}\
  \bibnamefont {Leotta}}}\ (\bibinfo  {publisher} {Springer},\ \bibinfo
  {address} {Boston, MA},\ \bibinfo {year} {1987})\ p.\ \bibinfo {pages}
  {195}\BibitemShut {NoStop}%
\bibitem [{\citenamefont {Noever}(1995)}]{noever1995}%
  \BibitemOpen
  \bibfield  {author} {\bibinfo {author} {\bibfnamefont {D.~A.}\ \bibnamefont
  {Noever}},\ }\bibfield  {title} {\bibinfo {title} {Polarized nuclei in a
  simple mirror fusion reactor},\ }\href
  {https://doi.org/10.13182/FST95-A30352} {\bibfield  {journal} {\bibinfo
  {journal} {Fusion Technol.}\ }\textbf {\bibinfo {volume} {27}},\ \bibinfo
  {pages} {86} (\bibinfo {year} {1995})}\BibitemShut {NoStop}%
\bibitem [{\citenamefont {Frank}\ \emph {et~al.}(2025)\citenamefont {Frank},
  \citenamefont {Viola}, \citenamefont {Petrov}, \citenamefont {Anderson},
  \citenamefont {Bindl} \emph {et~al.}}]{frank2025}%
  \BibitemOpen
  \bibfield  {author} {\bibinfo {author} {\bibfnamefont {S.}~\bibnamefont
  {Frank}}, \bibinfo {author} {\bibfnamefont {J.}~\bibnamefont {Viola}},
  \bibinfo {author} {\bibfnamefont {Y.}~\bibnamefont {Petrov}}, \bibinfo
  {author} {\bibfnamefont {J.}~\bibnamefont {Anderson}}, \bibinfo {author}
  {\bibfnamefont {D.}~\bibnamefont {Bindl}}, \emph {et~al.},\ }\bibfield
  {title} {\bibinfo {title} {Confinement performance predictions for a high
  field axisymmetric tandem mirror},\ }\href
  {https://doi.org/10.1017/S002237782510055X} {\bibfield  {journal} {\bibinfo
  {journal} {J. Plasma Phys.}\ }\textbf {\bibinfo {volume} {91}},\ \bibinfo
  {pages} {E110} (\bibinfo {year} {2025})}\BibitemShut {NoStop}%
\bibitem [{\citenamefont {Frank}\ \emph {et~al.}(2026)\citenamefont {Frank},
  \citenamefont {Viola}, \citenamefont {Petrov}, \citenamefont {Anderson},
  \citenamefont {Bindl} \emph {et~al.}}]{frank2026}%
  \BibitemOpen
  \bibfield  {author} {\bibinfo {author} {\bibfnamefont {S.}~\bibnamefont
  {Frank}}, \bibinfo {author} {\bibfnamefont {J.}~\bibnamefont {Viola}},
  \bibinfo {author} {\bibfnamefont {Y.}~\bibnamefont {Petrov}}, \bibinfo
  {author} {\bibfnamefont {J.}~\bibnamefont {Anderson}}, \bibinfo {author}
  {\bibfnamefont {D.}~\bibnamefont {Bindl}}, \emph {et~al.},\ }\bibfield
  {title} {\bibinfo {title} {Confinement performance predictions for a high
  field axisymmetric tandem mirror: corrigendum},\ }\href
  {https://doi.org/10.1017/S0022377826102037} {\bibfield  {journal} {\bibinfo
  {journal} {J. Plasma Phys.}\ }\textbf {\bibinfo {volume} {92}},\ \bibinfo
  {pages} {E94} (\bibinfo {year} {2026})}\BibitemShut {NoStop}%
\bibitem [{\citenamefont {Post}(1987)}]{post1987}%
  \BibitemOpen
  \bibfield  {author} {\bibinfo {author} {\bibfnamefont {R.~F.}\ \bibnamefont
  {Post}},\ }\bibfield  {title} {\bibinfo {title} {The magnetic mirror approach
  to fusion},\ }\href {https://doi.org/10.1088/0029-5515/27/10/001} {\bibfield
  {journal} {\bibinfo  {journal} {Nucl. Fusion}\ }\textbf {\bibinfo {volume}
  {27}},\ \bibinfo {pages} {1579} (\bibinfo {year} {1987})}\BibitemShut
  {NoStop}%
\end{thebibliography}%
\end{document}